\documentclass[letterpaper,oneside,final,onecolumn]{article}
\usepackage{graphicx}
\usepackage{amsmath}
\usepackage{amsfonts}
\usepackage{amssymb}

\begin{document}

\begin{center}
{\LARGE HIERARCHIC THEORY OF CONDENSED MATTER :}

\medskip

{\LARGE New state equation \&}{\large  }{\LARGE Interrelation between
mesoscopic and macroscopic properties}

\bigskip

\textbf{Alex Kaivarainen}

\textbf{\bigskip}

\textbf{JBL, University of Turku, FIN-20520, Turku, Finland\bigskip}

\textbf{URL: \thinspace http://www.karelia.ru/\symbol{126}alexk}

\textbf{H2o@karelia.ru}

\medskip
\end{center}

\medskip

\begin{quotation}
\textbf{Materials, presented in this original article are based on:}

\smallskip

\textbf{[1]. Book by A. Kaivarainen: Hierarchic Concept of Matter and Field.
Water, biosystems and elementary particles. New York, 1995 and two articles:}

\textbf{[2]. \thinspace New Hierarchic Theory of Matter General for Liquids
and Solids: }

\textbf{dynamics, thermodynamics and mesoscopic structure of water and ice }

\textbf{(see: http://arXiv.org/abs/physics/0003044\ and
http://www.karelia.ru/\symbol{126}alexk [New articles]);}

\textbf{[3].} \textbf{Hierarchic Concept of Condensed Matter and its
Interaction with Light: New Theories of Light Refraction, Brillouin
Scattering\ and M\"{o}ssbauer effect (see URL:
http://www.karelia.ru/\symbol{126}alexk [New articles]). }

\smallskip

\textbf{Computerized verification of described here new theories are presented
on examples of WATER and ICE, using special computer program (copyright, 1997,
A. Kaivarainen).}

\bigskip

\textbf{CONTENTS}

\smallskip

\textbf{Summary of Hierarchic Theory of Matter and Field}

\textbf{1 The state equation for real gas}

\textbf{2 New state equation for condensed matter}

\textbf{3 Vapor pressure}

\textbf{4 Surface tension}

\textbf{5 Mesoscopic theory of thermal conductivity}

\textbf{6 Mesoscopic theory of viscosity for liquids and solids}

\textbf{7 Brownian diffusion}

\textbf{8 Self-diffusion in liquids and solids}

\textbf{9 Mesoscopic approach to proton conductivity in water, ice and other
systems \ containing hydrogen bonds}

\textbf{10 Regulation of pH and shining of water by electromagnetic and
acoustic fields}

\newpage
\end{quotation}

\begin{center}
{\large Summary of }

{\large Hierarchic Theory of Condensed Matter (http://arXiv.org/abs/physics/0003044)}
\end{center}

{\large \smallskip}

\textbf{\ A basically new hierarchic quantitative theory, general for solids
and liquids, has been developed.}

\textbf{It is assumed, that unharmonic oscillations of particles in any
condensed matter lead to emergence of three-dimensional (3D) superposition of
standing de Broglie waves of molecules, electromagnetic and acoustic waves.
Consequently, any condensed matter could be considered as a gas of 3D standing
waves of corresponding nature. Our approach unifies and develops strongly the
Einstein's and Debye's models.}

\ \textbf{Collective excitations, like 3D standing de Broglie waves of
molecules, representing at certain conditions the mesoscopic molecular Bose
condensate, were analyzed, as a background of hierarchic model of condensed matter.}

\smallskip

\textbf{The most probable de Broglie wave (wave B) length is determined by the
ratio of Plank constant to the most probable impulse of molecules, or by ratio
of its most probable phase velocity to frequency. The waves B are related to
molecular translations (tr) and librations (lb).}

As far the quantum dynamics of condensed matter does not follow in general
case the classical Maxwell-Boltzmann distribution, the real most probable de
Broglie wave length can exceed the classical thermal de Broglie wave length
and the distance between centers of molecules many times.

\textit{This makes possible the atomic and molecular Bose condensation in
solids and liquids at temperatures, below boiling point. It is one of the most
important results of new theory, confirmed by computer simulations on examples
of water and ice.}

\smallskip

\textbf{Four strongly interrelated }new types of quasiparticles (collective
excitations) were introduced in our hierarchic model:

1.~\textit{Effectons (tr and lb)}, existing in "acoustic" (a) and "optic" (b)
states represent the coherent clusters in general case\textbf{; }

2.~\textit{Convertons}, corresponding to interconversions between \textit{tr
}and \textit{lb }types of the effectons (flickering clusters);

3.~\textit{Transitons} are the intermediate $\left[  a\rightleftharpoons
b\right]  $ transition states of the \textit{tr} and \textit{lb} effectons;

4.~\textit{Deformons} are the 3D superposition of IR electromagnetic or
acoustic waves, activated by \textit{transitons }and \textit{convertons. }\smallskip

\smallskip

\ \textbf{Primary effectons }(\textit{tr and lb) }are formed by 3D
superposition of the \textbf{most probable standing de Broglie waves }of the
oscillating ions, atoms or molecules. The volume of effectons (tr and lb) may
contain from less than one, to tens and even thousands of molecules. The first
condition means validity of \textbf{classical }approximation in description of
the subsystems of the effectons. The second one points to \textbf{quantum
properties} \textbf{of coherent clusters due to molecular Bose condensation}%
\textit{. }

\ The liquids are semiclassical systems because their primary (tr) effectons
contain less than one molecule and primary (lb) effectons - more than one
molecule. \textit{The solids are quantum systems totally because both kind of
their primary effectons (tr and lb) are molecular Bose condensates.}%
\textbf{\ These consequences of our theory are confirmed by computer
calculations. }

\ The 1st order $\left[  gas\rightarrow\,liquid\right]  $ transition is
accompanied by strong decreasing of rotational (librational) degrees of
freedom due to emergence of primary (lb) effectons and $\left[
liquid\rightarrow\,solid\right]  $ transition - by decreasing of translational
degrees of freedom due to mesoscopic Bose-condensation in form of primary (tr) effectons.

\ \textbf{In the general case the effecton can be approximated by
parallelepiped with edges corresponding to de Broglie waves length in three
selected directions (1, 2, 3), related to the symmetry of the molecular
dynamics. In the case of isotropic molecular motion the effectons' shape may
be approximated by cube. The edge-length of primary effectons (tr and lb) can
be considered as the ''parameter of order''.}\smallskip

The in-phase oscillations of molecules in the effectons correspond to the
effecton's (a) - \textit{acoustic }state and the counterphase oscillations
correspond to their (b) - \textit{optic }state. States (a) and (b) of the
effectons differ in potential energy only, however, their kinetic energies,
impulses and spatial dimensions - are the same. The \textit{b}-state of the
effectons has a common feature with \textbf{Fr\"{o}lich's polar mode. }

\smallskip

\textbf{The }$(a\rightarrow b)$\textbf{\ or }$(b\rightarrow a)$%
\textbf{\ transition states of the primary effectons (tr and lb), defined
as\ primary transitons, are accompanied by a change in molecule polarizability
and dipole moment without density fluctuations. At this case they lead to
absorption or radiation of IR photons, respectively. Superposition
(interception) of three internal standing IR photons of different directions
(1,2,3) - forms primary electromagnetic deformons (tr and lb).}

\ On the other hand, the [lb$\rightleftharpoons\,$tr] \textit{convertons }and
\textit{secondary transitons} are accompanied by the density fluctuations,
leading to \textit{absorption or radiation of phonons}.

\textit{Superposition, resulting from interception} of standing phonons in
three directions (1,2,3) is termed: \textbf{secondary acoustic deformons (tr
and lb). }

\smallskip

\ \textit{Correlated collective excitations }of primary and secondary
effectons and deformons (tr and lb)\textbf{, }localized in the volume of
primary \textit{tr }and \textit{lb electromagnetic }deformons\textbf{, }lead
to origination of \textbf{macroeffectons, macrotransitons}\textit{\ }and
\textbf{macrodeformons }(tr and lb respectively)\textbf{. }

\ \textit{Correlated simultaneous excitations of \thinspace tr and lb}
\textit{macroeffectons }in the volume of superimposed \textit{tr }and
\textit{lb }electromagnetic deformons lead to origination of
\textbf{supereffectons. }

\ In turn, the coherent excitation of \textit{both: tr }and \textit{lb
macrodeformons and macroconvertons }in the same volume means creation of
\textbf{superdeformons.} Superdeformons are the biggest (cavitational)
fluctuations, leading to microbubbles in liquids and to local defects in solids.

\smallskip

\ \textbf{Total number of quasiparticles of condensed matter equal to 4!=24,
reflects all of possible combinations of the four basic ones [1-4], introduced
above. This set of collective excitations in the form of ''gas'' of 3D
standing waves of three types: de Broglie, acoustic and electromagnetic - is
shown to be able to explain virtually all the properties of all condensed matter.}

\ \textit{The important positive feature of our hierarchic model of matter is
that it does not need the semi-empiric intermolecular potentials for
calculations, which are unavoidable in existing theories of many body systems.
The potential energy of intermolecular interaction is involved indirectly in
dimensions and stability of quasiparticles, introduced in our model.}

{\large \ The main formulae of theory are the same for liquids and solids and
include following experimental parameters, which take into account their
different properties:}

$\left[  1\right]  $\textbf{- Positions of (tr) and (lb) bands in oscillatory spectra;}

$\left[  2\right]  $\textbf{- Sound velocity; }$\,$

$\left[  3\right]  $\textbf{- Density; }

$\left[  4\right]  $\textbf{- Refraction index (extrapolated to the infinitive
wave length of photon}$)$\textbf{.}

\smallskip

\textit{\ The knowledge of these four basic parameters at the same temperature
and pressure makes it possible using our computer program, to evaluate more
than 300 important characteristics of any condensed matter. Among them are
such as: total internal energy, kinetic and potential energies, heat capacity
and thermal conductivity, surface tension, vapor pressure, viscosity,
coefficient of self-diffusion, osmotic pressure, solvent activity, etc. Most
of calculated parameters are hidden, i.e. inaccessible to direct experimental measurement.}

\ The new interpretation and evaluation of Brillouin light scattering and
M\"{o}ssbauer effect parameters may also be done on the basis of hierarchic
theory. Mesoscopic scenarios of turbulence, superconductivity and superfluity
are elaborated.

\ Some original aspects of water in organization and large-scale dynamics of
biosystems - such as proteins, DNA, microtubules, membranes and regulative
role of water in cytoplasm, cancer development, quantum neurodynamics, etc.
have been analyzed in the framework of Hierarchic theory.

\medskip

\textbf{Computerized verification of our Hierarchic theory of matter on
examples of water and ice is performed, using special computer program:
Comprehensive Analyzer of Matter Properties (CAMP, copyright, 1997,
Kaivarainen). The new optoacoustic device, based on this program, with
possibilities much wider, than that of IR, Raman and Brillouin spectrometers,
has been proposed (see URL:\thinspace\thinspace
http://www.karelia.ru/\symbol{126}alexk \ \ [CAMP]).}

\smallskip

\textbf{This is the first theory able to predict all known experimental
temperature anomalies for water and ice. The conformity between theory and
experiment is very good even without any adjustable parameters. }

\textbf{The hierarchic concept creates a bridge between micro- and macro-
phenomena, dynamics and thermodynamics, liquids and solids in terms of quantum physics.}

\bigskip

\begin{center}
********************************************************************
\end{center}

{\Large \medskip}

\begin{center}
{\Large 1. The state equation for real gas}
\end{center}

\smallskip

The Clapeyrone-Mendeleyev equation sets the relationship between pressure
$(P)$, volume $(V)$ and temperature ($T$) values for the ideal gas containing
$N_{0}$ molecules (one mole):%

\begin{equation}
PV=N_{0}kT=RT\tag{1}%
\end{equation}
In the real gases interactions between the molecules and their sizes should be
taken into account. It can be achieved by entering the corresponding
amendments into the left part, to the right or to the both parts of eq. (1).

It was Van der Waals who choused the first way more than a hundred years ago
and derived the equation:%

\begin{equation}
\left(  P+{\frac{a}{V^{2}}}\right)  \left(
\begin{array}
[c]{c}%
V-b
\end{array}
\right)  =RT\tag{2}%
\end{equation}
where the attraction forces are accounted for by the amending term $(a/V^{2})
$, while the repulsion forces and the effects of the excluded volume accounted
for the term (b).

Equation (2) correctly describes changes in P,V and T related to liquid-gas
transitions on the qualitative level. However, the quantitative analysis by
means of (2) is approximate and needs the fitting parameters. The parameters
(a) and (b) are not constant for the given substance and depend on
temperature. Hence, the Van der Waals equation is only some approximation
describing the state of a real gas.

We propose a way to modify the right part of eq.(1), substituting it for the
part of the kinetic energy (T) of 1 mole of the substance (eq.4.31 in [1, 2])
in real gas phase formed only by secondary effectons and deformons with
nonzero impulse, affecting the pressure:%

\[
PV={\frac23}\bar T_{\text{kin}}{={\frac23}V_{0}\frac1Z\cdot}\underset
{tr,lb}{\sum}\left[  \bar n_{ef}{\frac{\sum_{1}^{3}\left(  \bar E_{1,2,3}%
^{a}\right)  ^{2}}{2m\left(  \overline{v}_{ph}^{a}\right)  ^{2}}}\left(  \bar
P_{ef}^{a}+\bar P_{ef}^{b}\right)  +\right.
\]%

\begin{equation}
+\left.  \bar{n}_{d}{\frac{\sum_{1}^{3}\left(  \bar{E}_{d}^{1,2,3}\right)
^{2}}{2m\left(  v_{s}\right)  ^{2}}}\bar{P}_{d}\right]  _{tr,lb}\tag{3}%
\end{equation}
\textbf{The contribution to pressure caused by primary quasiparticles as
Bose-condensate with the zero resulting impulse is equal to zero also.}

\textbf{It is assumed when using such approach that for real gases the model
of a system of weakly interacted oscillator pairs is valid. The validity of
such an approach for water is confirmed by available experimental data
indicating the presence of dimers, trimers and larger }$H_{2}O$%
\textbf{\ clusters in the water vapor (Eisenberg and Kauzmann, 1975).}

\textbf{Water vapor has an intensive band in oscillatory spectra at }%
$\tilde{\nu}=200cm^{-1}$\textbf{. Possibly, it is this band that characterizes
the frequencies of quantum beats between ''acoustic'' (a) and ''optic'' (b)
translational oscillations in pairs of molecules and small clusters. The
frequencies of librational collective modes in vapor are absent.}

\textbf{The energies of primary gas quasiparticles }$(h\nu_{a}$%
\textbf{\ \ and\ }$h\nu_{b})$\textbf{\ can be calculated on the basis of the
formulae used for a liquid (Chapter 4 of [1] or [2]).}

\textbf{However, to calculate the energies of secondary quasiparticles in
(\={a}) and (\={b}) states the Bose-Einstein distribution must be used for the
case when the temperature is higher than the Bose-condensation temperature
}$(T>T_{0})$\textbf{\ and the chemical potential is not equal to zero }%
$(\mu<0)$\textbf{. According to this distribution:}%

\begin{equation}%
\begin{array}
[c]{l}%
\left\{  \bar{E}^{a}=h\bar{\nu}^{a}={\frac{h\nu^{a}}{\exp\left(  {\frac
{h\nu^{a}-\mu}{kT}}\right)  -1}}\right\}  _{tr,lb}\\
\left\{  \bar{E}^{b}=h\bar{\nu}^{b}={\frac{h\nu^{b}}{\exp\left(  {\frac
{h\nu^{b}-\mu}{kT}}\right)  -1}}\right\}  _{tr,lb}%
\end{array}
\tag{4}%
\end{equation}
\textbf{\medskip The kinetic energies of effectons }$(\bar{a})_{tr,lb}%
$\textbf{\ and }$(\bar{b})_{tr,lb}$\textbf{\ states are equal, only the
potential energies differ as in the case of condensed matter.}

\textbf{All other parameters in basic equation (3) can be calculated as
previously described [1, 2].}

\textbf{\medskip}

\begin{center}
{\Large 2. New state equation for condensed matter}
\end{center}

\smallskip

Using our eq.(4.3 from [1,2]) for the total internal energy of condensed
matter $(U_{\text{tot}})$, we can present state equation in a more general
form than (3).

For this end we introduce the notions of \textit{internal pressure
}$(P_{\text{in}})$, including \textbf{all type of interactions} between
particles of matter and excluded molar volume $(V_{\text{exc}})$:%

\begin{equation}
V_{\text{exc}}={\frac{4}{3}}\pi\alpha^{*}N_{0}=V_{0}\left(  {\frac{n^{2}%
-1}{n^{2}}}\right) \tag{5}%
\end{equation}
where $\alpha^{*}$ is the acting polarizability of molecules in condensed
matter $($see Part 1 of [3]$);\;N_{0}$ is Avogadro number, and $V_{0}$ is
molar volume.

\textbf{The general state equation }can be expressed in the following form:%

\begin{equation}
P_{\text{tot}}V_{fr}=(P_{\text{ext}}+P_{\text{in}})(V_{0}-V_{\text{exc}%
})=U_{ef}\tag{6}%
\end{equation}
where:\ $U_{ef}=U_{\text{tot}}(1+V/T_{\text{kin}}^{t})=U_{\text{tot}}%
^{2}/T_{\text{kin}}\;\;$ is the effective internal energy and:%

\[
(1+V/T_{\text{kin}})=U_{\text{tot}}/T_{\text{kin}}=S^{-1}
\]
is the reciprocal value of the total structural factor
$(eq.2.46a\,\,of\,\,[1]);\;P_{\text{tot}}=P_{\text{ext}}+P_{\text{in}}\;$ is
total pressure, $P_{\text{ext}}$ and $P_{\text{in}}$ are external and internal
pressures; $\;V_{fr}=V_{0}-V_{\text{exc}}=V_{0}/n^{2}\;($see eq.5) is a free
molar volume;\ $U_{\text{tot}}=V+T_{\text{kin}}$ is the total internal energy,
V and $T_{\text{kin}}$ are total potential and kinetic energies of one mole of matter.

For the limit case of ideal gas, when $P_{\text{in}}=0;\;V_{\text{exc}}=0$;
and the potential energy $\,V=0$, we get from (6) the Clapeyrone - Mendeleyev
equation (see 1):%

\[
P_{\text{ext}}V_{0}=T_{\text{kin}}=RT
\]

One can use equation of state (6) for estimation of sum of \textit{all types
of internal matter interactions}, which determines the internal pressure
$P_{\text{in}}$:%

\begin{equation}
P_{\text{in}}={\frac{U_{ef}}{V_{fr}}}-P_{\text{ext}}={\frac{n^{2}%
U_{\text{tot}}^{2}}{V_{0}T_{\text{kin}}}}-P_{\text{ext}}\tag{7}%
\end{equation}

where: the molar free volume: $V_{fr}=V_{0}-V_{\text{exc}}=V_{0}/n^{2}$; \ 

and the effective total energy: $U_{ef}=U_{\text{tot}}^{2}/T_{\text{kin}%
}=U_{\text{tot}}/$S.

For solids and most of liquids with a good approximation: $P_{\text{in}}%
\gg[P_{\text{ext}}\sim1$ atm. $=10^{5}Pa]$. Then from (7) we have:%

\begin{equation}
P_{\text{in}}\cong{\frac{n^{2}U_{\text{tot}}}{V_{0}S}}={\frac{n^{2}}{V_{0}}%
}\cdot U_{\text{tot}}\left(  1+{\frac{V}{T_{\text{kin}}}}\right) \tag{8}%
\end{equation}
where $S=T_{\text{kin}}/U_{\text{tot}}$ is a total structural factor;
$T_{\text{kin}}$ and V are total kinetic and potential energies, respectively.

For example for 1 mole of water under standard conditions we obtain: $\;\;\;\;\;\;\;\,\,\,$

$V_{\text{exc}}=8.4cm^{3};V_{fr}=9.6cm^{3};\;\;V_{0}=V_{\text{exc}}%
+V_{fr}=18cm^{3};$

$P_{\text{in}}\cong380000$ atm. $=3.8\cdot10^{10}Pa$ \ (1 atm. =10$^{5}Pa). $

The parameters such as sound velocity, molar volume, and the positions of
translational and librational bands in oscillatory spectra that determine
$U_{ef}(4.3)$ depend on external pressure and temperature.

The results of computer calculations of $P_{\text{in}}\;(eq.7)$ for ice and
water are presented on Fig. 1 a,b.

Polarizability and, consequently, free volume $(V_{fr})$ and $P_{\text{in}}$
in (6) depend on energy of external electromagnetic fields (see Part 1 of [3].

\begin{center}%
\begin{center}
\includegraphics[
height=2.1223in,
width=4.9658in
]%
{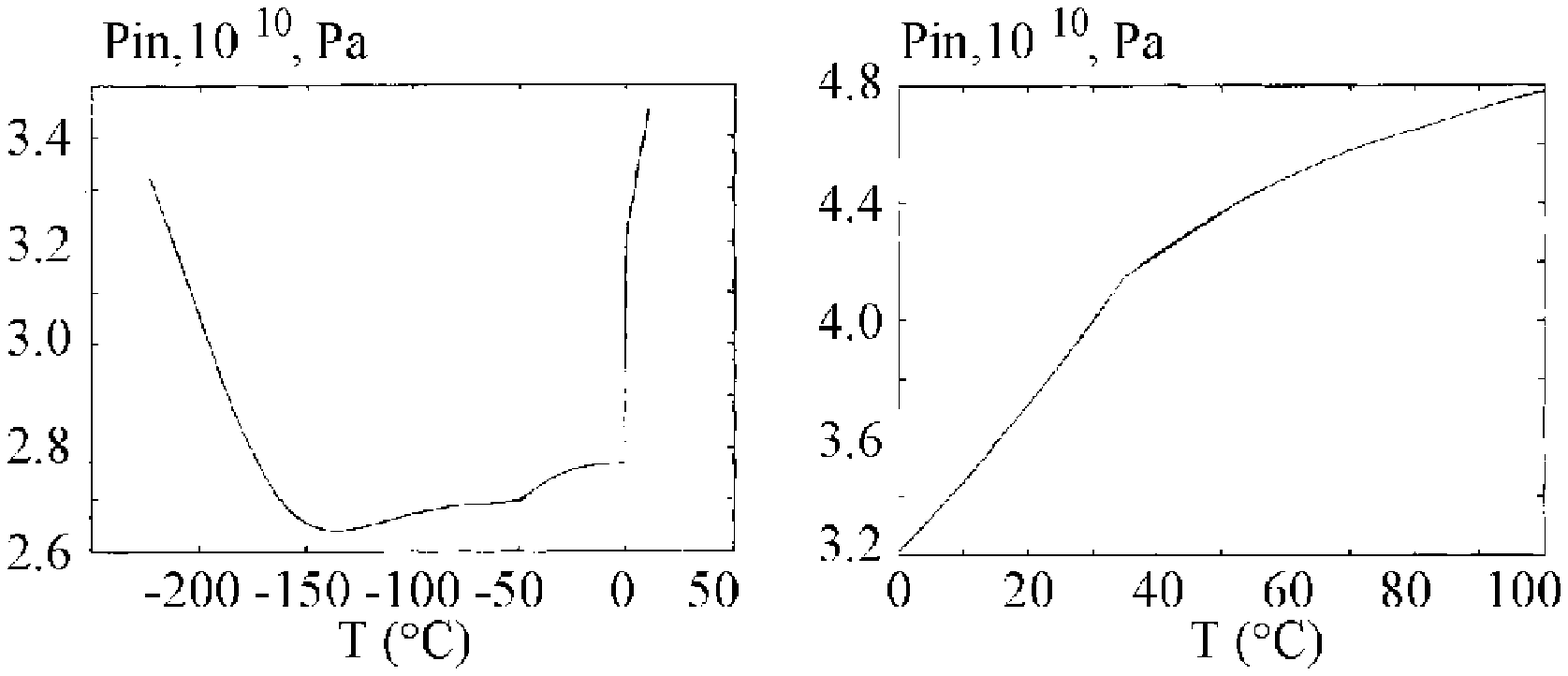}%
\end{center}
\medskip
\end{center}
\nopagebreak
\begin{quotation}
\textbf{Fig.~1. }(a) Theoretical temperature dependence of internal pressure
$(P_{\text{in}})$ in ice including the point of [ice $\Leftrightarrow$ water]
phase transition; (b) Theoretical temperature dependence of internal pressure
$(P_{\text{in}})$ in water. Computer calculations were performed using eq. (7).
\end{quotation}

\medskip

The minima of $P_{\text{in}}(T)$ for ice at $-140^{0}$ and $-50^{0}C$ in
accordance with eq.(9) correspond to the most stable structure of this matter,
related to temperature transition. In water some kind of transition appears at
$35^{0}C$, near physiological temperature.

There may exist conditions when the derivatives of internal pressure
P$_{\text{in}}$ are equal to zero:%

\begin{equation}
(a):\left(  {\frac{\partial P_{\text{in}}}{\partial P_{\text{ext}}}}\right)
_{T}=0\text{ \thinspace\ and\ \ \ }(b):\left(  {\frac{\partial P_{\text{in}}%
}{\partial T}}\right)  _{P_{\text{ext}}}=0\tag{9}%
\end{equation}
This condition corresponds to the \textbf{minima of potential energy, i.e. to
the most stable structure of given matter. }In a general case there may be a
few metastable states when conditions (9) are fulfilled.

Equation of state (7) may be useful for the study of mechanical properties of
condensed matter and their change under different influences.

Differentiation of (6) by external pressure gives us at T = \textit{const: }%

\begin{equation}
V_{fr}+{\frac{\partial P_{fr}}{\partial P_{\text{ext}}}}(P_{\text{ex}%
}+P_{\text{in}})+V_{fr}{\frac{\partial P_{\text{in}}}{\partial P_{\text{ext}}%
}}={\frac{\partial P_{ef}}{\partial P_{\text{ext}}}}\tag{10}%
\end{equation}
Dividing the left and right part of (10) by free volume $V_{fr}$ we obtain:%

\begin{equation}
\left(  {\frac{\partial P_{\text{in}}}{\partial P_{\text{ext}}}}\right)
_{T}=\left(  {\frac{\partial P_{ef}}{\partial P_{\text{ext}}}}\right)
_{T}-\left[
\begin{array}
[c]{c}%
1+\beta_{T}(P_{\text{ext}}+P_{\text{in}})
\end{array}
\right]  _{T}\tag{11}%
\end{equation}
where:\ $\beta_{T}=-(\partial V_{fr}/\partial P_{\text{ext}})/V_{fr}$ is
isothermal compressibility. From (9) and (11) we derive condition for the
\textit{maximum stability }of matter structure:%

\begin{equation}
\left(  {\frac{\partial P_{ef}}{\partial P_{\text{ext}}}}\right)  _{T}%
=1+\beta_{T}^{0}P_{\text{tot}}^{\text{opt}}\tag{12}%
\end{equation}

where: $P_{\text{tot}}^{\text{opt}}=P_{\text{ext}}+P_{\text{in}}^{\text{opt}}
$ is the ''optimum'' total pressure.

The derivative of (6) by temperature gives us at $P_{\text{ext}}=$
\textit{const}:%

\begin{equation}
P_{\text{tot}}\left(  {\frac{\partial V_{fr}}{\partial T}}\right)
_{P_{\text{ext}}}+V_{fr}\left(  {\frac{\partial P_{\text{in}}}{\partial T}%
}\right)  _{P_{\text{ext}}}=\left(  {\frac{\partial U_{ef}}{\partial T}%
}\right)  _{P_{\text{ext}}}=C_{V}\tag{13}%
\end{equation}
where%

\begin{align}
\left(  {\frac{\partial V_{fr}}{\partial T}}\right)  _{P_{\text{ext}}}  &
=\left(  {\frac{\partial V_{0}}{\partial T}}\right)  _{P_{\text{ext}}}%
-{\frac{4}{3}}\pi N_{0}\left(  {\frac{\partial\alpha^{*}}{\partial T}}\right)
_{P_{\text{ext}}}\text{ }\tag{14}\\
\text{and\ \ \ \ }\left(  {\frac{\partial V_{\text{tot}}}{\partial T}}\right)
_{P_{\text{ext}}}  & ={\frac{\partial P_{\text{in}}}{\partial T}}\tag{14a}%
\end{align}
\medskip From our mesoscopic theory of refraction index (Part 1 of [3]) the
acting polarizability $\alpha^{*}$ is:%

\begin{equation}
\alpha^{*}=\frac{\left(  {\frac{n^{2}-1}{n^{2}}}\right)  }{{\frac{4}{3}}%
\pi{\frac{N_{0}}{V_{0}}}}\tag{15}%
\end{equation}
When condition (9b) is fulfilled, we obtain for optimum internal pressure
$(P_{\text{in}}^{\text{opt}})$ from (13):%

\begin{equation}
P_{\text{in}}^{\text{opt}}=C_{V}/\left(  {\frac{\partial V_{fr}}{\partial T}%
}\right)  _{P_{\text{ext}}}-P_{\text{ext}}\tag{16}%
\end{equation}
or%

\begin{equation}
P_{\text{in}}^{\text{opt}}={\frac{C}{V_{fr}\gamma}}-P_{\text{ext}},\tag{17}%
\end{equation}

where%

\begin{equation}
\gamma=(\partial V_{fr}/\partial T)/V_{fr}\tag{18}%
\end{equation}
is the thermal expansion coefficient;

V$_{fr}$ is the total free volume in 1 mole of condensed matter:%

\begin{equation}
V_{fr}=V_{0}-V_{\text{exc}}=V_{0}/n^{2}\tag{19}%
\end{equation}
It is taken into account in (13) and (19) that%

\begin{equation}
(\partial V_{\text{exc}}/\partial T)\cong0\tag{20}%
\end{equation}
because, as has been shown earlier (Fig.25a of [1] and Part 1 of [3]),%

\[
\partial\alpha^{*}/\partial T\cong0
\]
Dividing the left and right parts of (13) by $P_{\text{tot}}V_{fr}=U_{ef}$, we
obtain for the heat expansion coefficient:%

\begin{equation}
\gamma={\frac{C_{V}}{U_{ef}}}-{\frac{1}{P_{\text{tot}}}}\left(  {\frac
{\partial P_{\text{in}}}{\partial T}}\right)  _{P_{\text{ext}}}\tag{21}%
\end{equation}
Under metastable states, when condition (9 b) is fulfilled,%

\begin{equation}
\gamma^{0}=C_{V}/U_{ef}\tag{22}%
\end{equation}

Putting (8) into (12), we obtain for isothermal compressibility of metastable
states corresponding to (9a) following formula:%

\begin{equation}
\beta_{T}^{0}={\frac{V_{0}T_{\text{kin}}}{n^{2}U_{\text{tot}}^{2}}}\left(
{\frac{\partial U_{ef}}{\partial P_{\text{ext}}}}-1\right) \tag{23}%
\end{equation}

It seems that our equation of state (7) may be used to study different types
of external influences (pressure, temperature, electromagnetic radiation,
deformation, etc.) on the thermodynamic and mechanic properties of solids and liquids.

\medskip

\begin{center}
{\Large 3. Vapor pressure}
\end{center}

\smallskip

When a liquid is incubated long enough in a closed vessel at constant
temperature, then an equilibrium between the liquid and vapor is attained.

At this moment, the number of molecules evaporated and condensed back to
liquid is equal. The same is true of the process of sublimation.

There is still no satisfactory quantitative theory for \textit{vapor pressure
}calculation.

We can suggest such a theory using our notion of \textit{superdeformons},
representing the biggest thermal fluctuations (see Table 1 and Introduction).
The basic idea is that the external equilibrium vapor pressure is related to
internal one $(P_{\text{in}}^{S})$ with coefficient determined by the
probability of cavitational fluctuations (superdeformons) in the
\textbf{surface layer} of liquids or solids.

In other words due to excitation of superdeformons with probability
$(P^{S}_{D})$, the internal pressure $(P^{S}_{\text{in}})$ in surface layers,
determined by the total contributions of all intramolecular interactions turns
to external one - vapor pressure $(P_{V})$. It is something like a compressed
spring energy realization due to trigger switching off.

For taking into account the difference between the surface and bulk internal
pressure $(P_{\text{in}})$ we introduce the semiempirical surface pressure
factor $(q^{S})$ as:%

\begin{equation}
P_{\text{in}}^{S}=q^{S}P_{\text{in}}-P_{\text{ext}}=q^{S}\cdot{\frac
{n^{2}U_{\text{tot}}}{V_{0}S}}-P_{\text{ext}}\tag{24}%
\end{equation}
where: P$_{\text{in}}$ corresponds to $eq.(7);\;\;S=T_{\text{kin}%
}/U_{\text{tot}}$ is a total structure factor.

The value of surface factor $(q^{S})$ for liquid and solid states is not the same:%

\vspace{-3mm}

\begin{equation}
q_{\text{liq}}^{S}<q_{\text{sol}}^{S}\tag{25}%
\end{equation}

\ 

\vspace{-7mm}
\begin{center}%
\begin{center}
\includegraphics[
height=2.2001in,
width=4.9658in
]%
{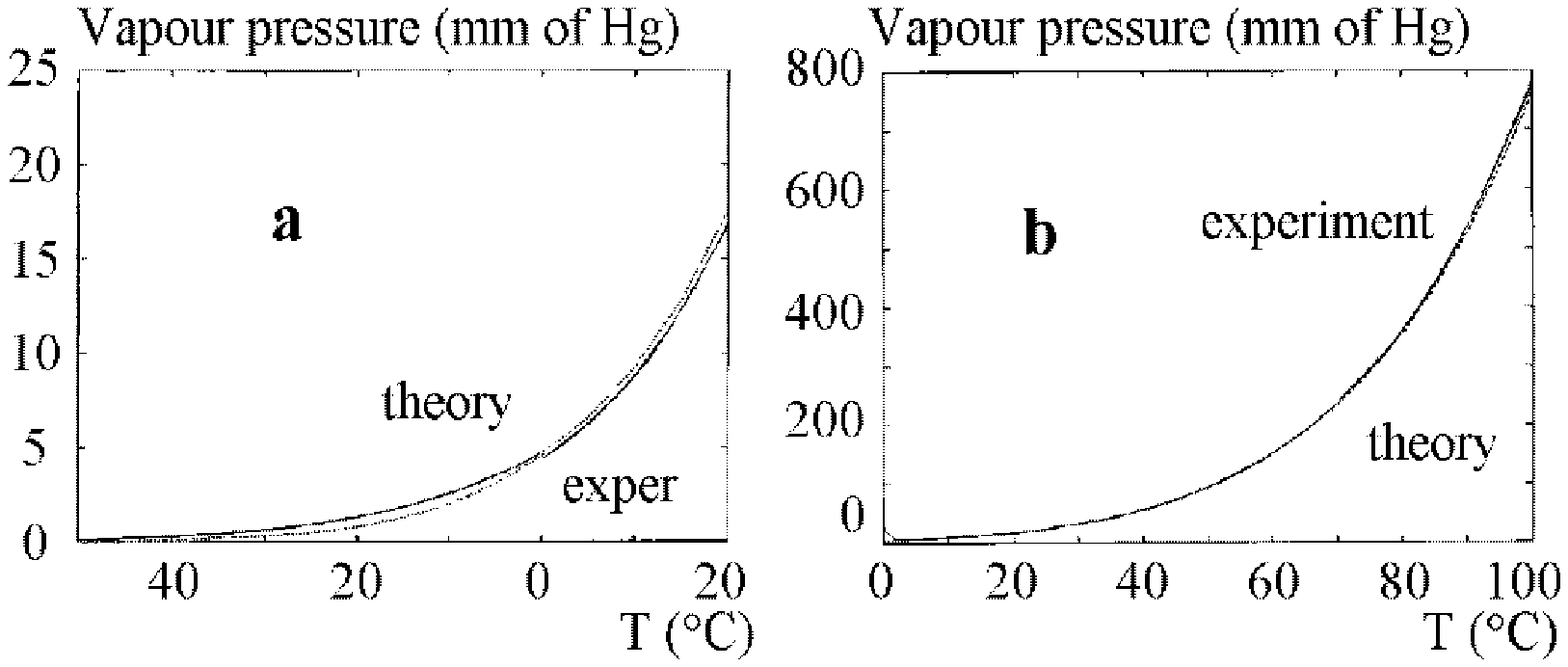}%
\end{center}
\medskip
\end{center}
\vspace{-10mm}
\begin{quotation}
\textbf{Fig.~2. }a) Theoretical $(-)$ and experimental ($\cdot\cdot$)
temperature dependences of vapor pressure $(P_{\text{vap}})$ for ice (a) and
water (b) including phase transition region. Computer calculations were
performed using eq. (26).
\end{quotation}

\medskip

Multiplying (24) to probability of superdeformons excitation we obtain for
vapor pressure, resulting from evaporation or sublimation, the following formulae:%

\begin{equation}
P_{vap}=P_{\text{in}}^{S}\cdot P_{D}^{S}=\left(  q^{S}{\frac{n^{2}%
U_{\text{tot}}^{2}}{V_{0}T_{\text{kin}}}}-P_{\text{ext}}\right)  \cdot
\exp\left(  -{\frac{E_{D}^{S}}{kT}}\right) \tag{26}%
\end{equation}

where:%

\begin{equation}
P_{D}^{S}=\exp\left(  -{\frac{E_{D}^{S}}{kT}}\right) \tag{27}%
\end{equation}
is a probability of superdeformons excitation (see eqs. 3.37, 3.32 and 3.33).

We can assume, that the difference in the surface and bulk internal pressure
is determined mainly by difference in total internal energy $(U_{\text{tot}})
$ but not in kinetic one $(T_{k})$. Then a pressure surface factor could be
presented as:
\[
q^{S}=\gamma^{2}=(U_{\text{in}}/U_{\text{tot}})^{2}
\]
\ where: $\gamma=U_{\text{tot}}^{S}/U_{\text{tot}}$ is the \textbf{surface
energy factor}, reflecting the ratio of surface and bulk total energy.

Theoretical calculated temperature dependences of vapor pressure, described by
(26) coincide very well with experimental ones for water at $q_{\text{liq}%
}^{S}=3.1\;(\gamma_{l}=1.76)$ and for ice at $q_{\text{sol}}^{S}%
=18\;(\gamma_{s}=4.24)\;($Fig. 2).

The almost five-times difference between q$_{\text{sol}}^{S}$
and\ $q_{\text{liq}}^{S}\,$ means that the \textit{surface} properties of ice
differ from \textit{bulk} ones much more than for liquid water.

\textbf{The surface factors }$q_{\text{liq}}^{S}$\textbf{\ and }%
$q_{\text{sol}}^{S} $\textbf{\ should be considered as a fit parameters. The
}$\;q^{S}=$\textbf{\ }$\gamma^{2}\,\,$\textbf{\ is the only one fit parameter
that was used in our hierarchic mesoscopic theory. Its calculation from the
known vapor pressure or surface tension can give an important information
itself. }

\medskip

\begin{center}
{\Large 4. Surface tension}
\end{center}

\smallskip

The resulting surface tension is introduced in our mesoscopic model as a sum:%

\begin{equation}
\sigma=(\sigma_{tr}+\sigma_{lb})\tag{28}%
\end{equation}
where: $\sigma_{tr}$ and $\sigma_{lb}$ are translational and librational
contributions to surface tension. Each of these components can be expressed
using our mesoscopic state equation (6, 7), taking into account the difference
between surface and bulk total energies $(q^{S})$, introduced in previous section:%

\begin{equation}
\sigma_{tr,lb}={\frac{1}{{\frac{1}{\pi}}(V_{ef})_{tr,lb}^{2/3}}}\left[
{\frac{q^{S}P_{\text{tot}}(P_{ef}V_{ef})_{tr,lb}-P_{\text{tot}}(P_{ef}%
V_{ef})_{tr,lb}}{(P_{ef}+P_{t})_{tr}+(P_{ef}+P_{t})_{lb}+(P_{\text{con}%
}+P_{\text{cMt}})}}\right] \tag{29}%
\end{equation}
where $(V_{ef})_{tr,lb}$ are volumes of primary tr and lib effectons, related
to their concentration $(n_{ef})_{tr,lb}$ as:%

\[
(V_{ef})_{tr,lb}=(1/n_{ef})_{tr,lb};
\]
\[
r_{tr,lb}={\frac{1}{\pi}}(V_{ef})_{tr,lb}^{2/3}
\]
is an effective radius of the primary translational and librational effectons,
localized on the surface of condensed matter; $q^{S}$ is the surface factor,
equal to that used in eq.(24\thinspace-26)$;\;\;\left[  P_{\text{tot}%
}=P_{\text{in}}+P_{\text{ext}}\right]  $ is a total pressure, corresponding to
eq.(6);$\;(P_{ef})_{tr,lb}$ is a total probability of primary effecton
excitations in the (a) and (b) states:%

\[
(P_{ef})_{tr}=(P_{ef}^{a}+P_{ef}^{b})_{tr}
\]
\[
(P_{ef})_{lb}=(P_{ef}^{a}+P_{ef}^{b})_{lb}
\]
$(P_{t})_{tr}$ and $(P_{t})_{lb}$ in (29) are the probabilities of
corresponding transiton excitation;

$P_{\text{con}}=P_{ac}+P_{bc}$ is the sum of probabilities of $\left[
\mathit{a}\right]  $\textit{\ }and $\left[  \mathit{b}\right]  $%
\textit{\ convertons;\ \thinspace}$P_{\text{cMt}}=P_{ac}\cdot P_{bc}\,$ is the
probability of Macroconverton (see Introduction and Chapter 4).

The eq. (29) contains the ratio:%

\begin{equation}
(V_{ef}/V_{ef}^{2/3})_{tr,lb}=l_{tr,lb}\tag{30}%
\end{equation}
where: $\;l_{tr}=(1/n_{ef})_{tr}^{1/3}$ \ and $\;l_{lb}=(1/n_{ef}%
)_{\text{lib}}^{1/3}$ \ are the length of the ribs of the primary
translational and librational effectons, approximated by cube.

Using (30) and (29) the resulting surface tension (28) can be presented as:%

\begin{equation}
\sigma=\sigma_{tr}+\sigma_{lb}=\pi{\frac{%
\begin{array}
[c]{c}%
P_{\text{tot}}(q^{S}-1)\cdot\left[
\begin{array}
[c]{c}%
(P_{ef})_{tr}l_{tr}+(P_{ef})l_{lb}%
\end{array}
\right]
\end{array}
}{(P_{ef}+P_{t})_{tr}+(P_{ef}+P_{t})_{lb}+(P_{\text{con}}+P_{\text{cMt}})}%
}\tag{31}%
\end{equation}
where translational component of surface tension is:%

\begin{equation}
\sigma_{tr}=\pi{\frac{P_{\text{tot}}(q^{s}-1)(P_{ef})_{tr}l_{tr}}%
{(P_{ef}+P_{t})_{tr}+(P_{ef}+P_{t})_{lb}+(P_{\text{con}}+P_{\text{cMt}})}%
}\tag{32}%
\end{equation}

and librational component of $\sigma$ is:%

\begin{equation}
\sigma_{lb}=\pi{\frac{P_{\text{tot}}(q^{S}-1)(P_{ef})_{lb}l_{lb}}%
{(P_{ef}+P_{t})_{lb}+(P_{ef}+P_{t})_{lb}+(P_{\text{con}}+P_{\text{cMt}})}%
}\tag{33}%
\end{equation}

Under the boiling condition when q$^{S}\rightarrow$ 1 as a result of
$\;(U_{\text{tot}}^{S}\rightarrow U_{\text{tot}})$, then $\sigma_{tr}%
,\;\sigma_{lb}\,$ and\thinspace$\sigma\,$ tends to zero. The maximum depth of
the surface layer, which determines the\thinspace$\sigma_{lb}\,\;$is equal to
the length of edge of cube $\left(  l_{lb}\right)  $, that approximates the
shape of primary \textbf{librational }effectons. It decreases from about 20
\AA\ at 0$^{0}C\,$ till about 2.5 \AA\ at 100$^{0}C\;($see$\,$Fig. 7b of [1]
or Fig. 4b of [2]). Monotonic decrease of $\left(  l_{lb}\right)  \,$with
temperature could be accompanied by nonmonotonic change of probabilities of
[lb/tr] convertons and macroconvertons excitations (see comments to Fig. 7a of
[1] or to Fig 4a of [2]). Consequently, the temperature dependence of surface
tension on temperature can display anomalies at definite temperatures. This
consequence of our theory is confirmed experimentally (Adamson, 1982;
Drost-Hansen and Lin Singleton, 1992).

The thickness of layer $\left(  l_{tr}\right)  $, responsible for contribution
of \textbf{translational} effectons in surface tension $\left(  \sigma
_{tr}\right)  \;$has the dimension of one molecule in all temperature interval
for liquid water.

The results of computer calculations of $\sigma\;$(eq.31) for water and
experimental data are presented at Fig.3.

\begin{center}%
\begin{center}
\includegraphics[
height=2.1223in,
width=4.9658in
]%
{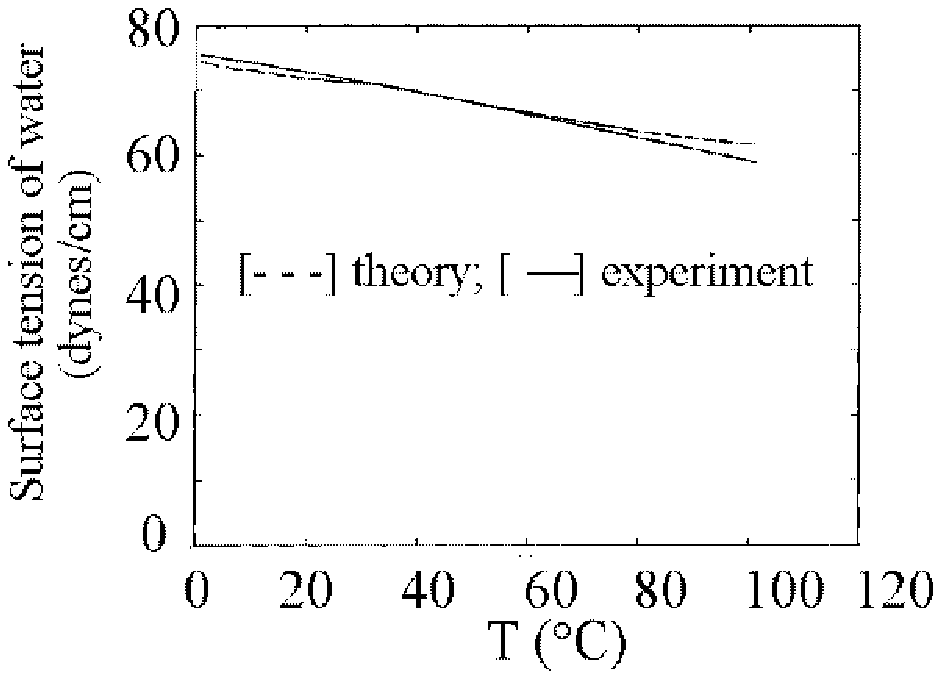}%
\end{center}
\medskip
\end{center}

\begin{quotation}
\textbf{Fig.~3. }Experimental $(^{\_\_\_})$ and theoretical (-\thinspace
-\thinspace-) temperature dependences of the surface tension for water,
calculated from eq.(31).
\end{quotation}

\medskip

\textbf{It is obvious, that the correspondence between theory and experiment
is very good, confirming in such a way the correctness of our model and
Hierarchic concept in general.\ }

\medskip

\begin{center}
{\Large 5. Mesoscopic theory of thermal conductivity}
\end{center}

\smallskip

\textbf{Thermal conductivity may be related to phonons, photons, free
electrons, holes and [electron-hole] pairs movement.}

\textbf{We will discuss here only the main type of thermal conductivity in
condensed matter, related to phonons.}

\textbf{The analogy with the known formula for thermal conductivity (}$\kappa
$\textbf{) in the framework of the kinetic theory for gas is used [4]:}%

\begin{equation}
\kappa={\frac{1}{3}}C_{v}v_{s}\Lambda\tag{34}%
\end{equation}
where C$_{v}$ is the heat capacity of condensed matter, $v_{s}$ is sound
velocity, characterizing the speed of phonon propagation in matter, and
$\Lambda$ is the average \textbf{length of free run} of phonons.

The value of $\Lambda$ depends on the scattering and dissipation of phonons at
other phonons and different types of defects. Usually decreasing temperature
increases $\Lambda$.

Different factors influencing a thermal equilibrium in the system of phonons
are discussed. Among them are the so called U- and N- processes describing the
types of phonon-phonon interaction. However, the traditional theories are
unable to calculate $\Lambda$ directly.

Mesoscopic theory introduce two contributions to thermal conductivity: related
to phonons, irradiated by secondary effectons and forming \textbf{secondary
}translational and librational deformons ($\kappa_{sd}$)$_{tr,lb}$ and to
phonons, irradiated by $\mathit{a}$\textit{\ }and $\mathit{b}$\textit{\ }%
convertons $[tr/lb]$, forming the convertons-induced deformons $(\kappa
_{cd})_{ac.bc}$:%

\begin{equation}
\kappa=(\kappa_{sd})_{tr,lb}+(\kappa_{cd})_{ac.bc}={\frac{1}{3}}C_{v}%
v_{s}[(\Lambda_{sd})_{tr,lb}+(\Lambda_{cd})_{ac,bc}]\tag{35}%
\end{equation}

where: \textbf{free runs }of secondary phonons (tr and lb) are represented as:%

\[
1/(\Lambda_{sd})_{tr,lb}=1/(\Lambda_{tr})+1/(\Lambda_{lb})=(\overline{\nu}%
_{d})_{tr}/v_{s}+(\overline{\nu}_{d})_{lb}/v_{s}
\]
consequently:
\begin{equation}
1/(\Lambda_{sd})_{tr,lb}=\frac{v_{s}}{(\overline{\nu}_{d})_{tr}+(\overline
{\nu}_{d})_{lb}}\tag{36}%
\end{equation}
\medskip and free runs of convertons-induced phonons:%

\[
1/(\Lambda_{cd})_{ac,bc}=1/(\Lambda_{ac})+1/(\Lambda_{bc})=(\nu_{ac}%
)/v_{s}+(\nu_{bc})/v_{s}
\]
\begin{equation}
\text{consequently: }(\Lambda_{sd})_{tr,lb}={\frac{v_{s}}{(\nu_{d})_{tr}%
+(\nu_{d})_{lb}}}\tag{37}%
\end{equation}
The heat capacity: $C_{V}=\partial U_{\text{tot}}/\partial T$ can be
calculated also from our theory (see Chapter 4 and 5).

\begin{center}%
\begin{center}
\includegraphics[
height=2.4708in,
width=5.7925in
]%
{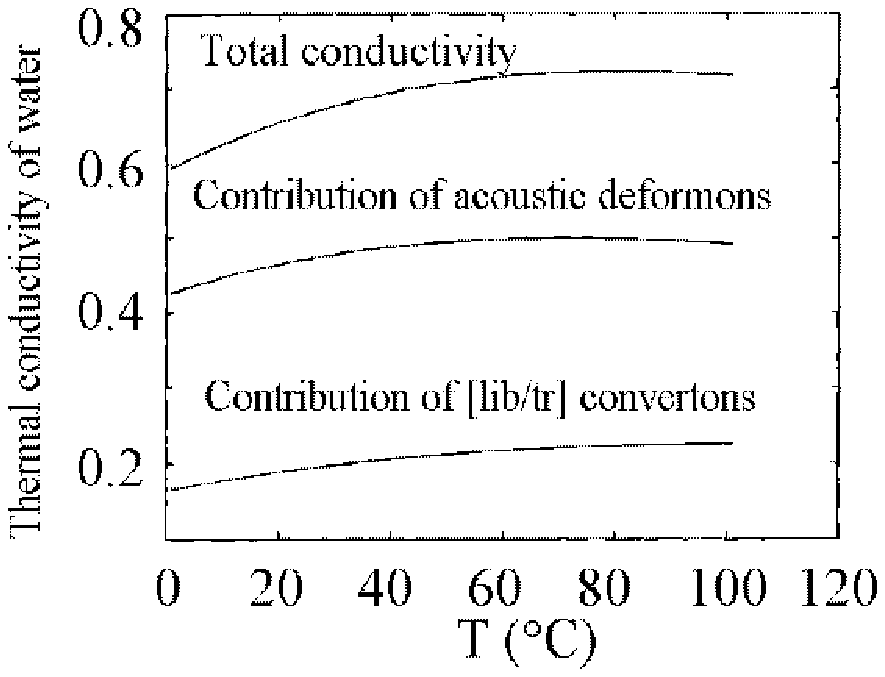}%
\end{center}
\medskip
\end{center}

\begin{quotation}
\textbf{Fig.~4. }Temperature dependences of total thermal conductivity for
water and contributions, related to acoustic deformons and $[lb/tr]$%
convertons. The dependences were calculated, using eq. (37).
\end{quotation}

\smallskip

Quantitative calculations show that formula (35), based on our mesoscopic
model, works well for water (Fig. 4).\textbf{\ It could be used for any other
condensed matter also if positions of translational and librational bands,
sound velocity and molar volume for this matter at the same temperature
interval are known.}

\textbf{The small difference between experimental and theoretical data can
reflect the contributions of non-phonon process in thermal conductivity,
related to macrodeformons, superdeformons and macroconvertons, i.e. big fluctuations.}

\medskip

\begin{center}
{\Large 6. Mesoscopic theory of viscosity for liquids and solids}
\end{center}

\smallskip

\textbf{The viscosity }is determined by the energy dissipation as a result of
medium (liquid or solid) structure deformation. Viscosity corresponding to the
shift deformation is named \textit{shear viscosity}. So- called \textit{bulk
viscosity }is related to deformation of volume parameters and corresponding
dissipation. These types of viscosity have not the same values and nature.

The statistical theory of irreversible process leads to the following
expression for shear viscosity (Prokhorov, 1988):%

\begin{equation}
\eta=\text{ }nkT\tau_{p}+(\mu_{\infty}\text{ }-\text{\thinspace\thinspace
}nkT)\tau_{q}\tag{38}%
\end{equation}
where n is the concentration of particles, $\mu_{\infty}$ is the modulus of
instant shift characterizing the instant elastic reaction of medium, $\tau
_{p}$ and $\tau_{q}$ are the relaxation times of impulses and coordinates, respectively.

However, eq.(38) is inconvenient for practical purposes due to difficulties in
determination of $\tau_{p},\tau_{q}$ and $\mu_{\infty}$.

Sometimes in a narrow temperature interval the empiric Ondrade equation is working:%

\begin{equation}
\eta=A(T)\cdot\exp(\beta/T)\tag{39}%
\end{equation}
A(T) is a function poorly dependent on temperature.

A good results in study the microviscosity problem were obtained by combining
the model of molecular rotational relaxation [5] and the Kramers equation (\AA
kesson et al., 1991). However, the using of the fit parameters was necessarily
in this case also.

\textbf{We present here our mesoscopic theory of viscosity. }To this end the
dissipation processes, related to ($A\rightleftharpoons B)_{tr.lb}$ cycles of
translational and librational macroeffectons and (a,b)-\textit{convertons
}excitations were used. The same approach was employed for elaboration of
mesoscopic theory of diffusion in condensed matter (see next section).

\textit{In contrast to liquid state, the viscosity of solids }is determined by
the biggest fluctuations: \textbf{supereffectons} and \textbf{superdeformons},
resulting from simultaneous excitations of translational and librational
macroeffectons and macrodeformons in the same volume.

The dissipation phenomena and ability of particles or molecules to diffusion
are related to the local fluctuations of the free volume $(\Delta
v_{f})_{tr,lb}$. According to mesoscopic theory, the fluctuations of free
volume and that of density occur in the almost macroscopic volumes of
translational and librational macrodeformons and in mesoscopic volumes of
\textbf{macroconvertons}, equal to volume of primary librational effecton at
the given conditions. Translational and librational types of macroeffectons
determine two types of viscosity, i.e. translational $(\eta_{tr})$ and
librational $(\eta_{lb})$ ones. They can be attributed to the bulk viscosity.
The contribution to viscosity, determined by \textit{(a and b)- convertons is
much more local and may be responsible for microviscosity and mesoviscosity.}

Let us start from calculation of the additional free volumes $(\Delta v_{f})$
originating from fluctuations of density, accompanied the translational and
librational macrodeformons (macrotransitons).

For 1 mole of condensed matter the following ratio between free volume and
concentration fluctuations is true:%

\begin{equation}
\left(  {\frac{\Delta v_{f}}{v_{f}}}\right)  _{tr,lb}=\left(  {\frac{\Delta
N_{0}}{N_{0}}}\right)  _{tr,lb}\tag{40}%
\end{equation}
where $N_{0}$ is the average number of molecules in 1 mole of matter%

\begin{equation}
\text{and\qquad}(\Delta N_{0})_{tr,lb}=N_{0}\left(  {\frac{P_{D}^{M}}{Z}%
}\right)  _{tr,lb}\tag{41}%
\end{equation}
is the number of molecules changing their concentration as a result of
translational and librational macrodeformons excitation.

The probability of translational and librational macroeffectons excitation
(see eqs. 3.23; 3.24):%

\begin{equation}
\left(  {\frac{P_{D}^{M}}{Z}}\right)  _{tr,lb}={\frac{1}{Z}}\exp\left(
-{\frac{\epsilon_{D}^{M}}{kT}}\right)  _{tr,lb}\tag{42}%
\end{equation}
where Z is the total partition function of the system (Chapter 4 of [1, 2]).

Putting (41) to (40) and dividing to Avogadro number $(N_{0})$, we obtain the
fluctuating free volume, reduced to 1 molecule of matter:%

\begin{equation}
\Delta v_{f}^{0}={\frac{\Delta v_{f}}{N_{0}}}=\left[  {\frac{v_{f}}{N_{0}}%
}\left(  {\frac{P_{D}^{M}}{Z}}\right)  \right]  _{tr,lb}\tag{43}%
\end{equation}
It has been shown above (eq.19) that the average value of free volume in 1
mole of matter is:%

\[
v_{f}=V_{0}/n^{2}
\]
Consequently, for reduced fluctuating (additional) volume we have:%

\begin{equation}
(\Delta v_{f}^{0})_{tr,lb}={\frac{V_{0}}{N_{0}n^{2}}}{\frac{1}{Z}}\exp\left(
-{\frac{\epsilon_{D}^{M}}{kT}}\right)  _{tr,lb}\tag{44}%
\end{equation}

Taking into account the dimensions of viscosity and its physical sense, it
should be proportional to the work (activation energy) of
fluctuation-dissipation, necessary for creating the unit of additional free
volume: $(E_{D}^{M}/\Delta v_{f}^{0})$, and the period of
($A\rightleftharpoons B)_{tr.lb}$ cycles of translational and librational
macroeffectons $\tau_{A\rightleftharpoons B},\,$determined by the life-times
of all intermediate states (eq.46).

\textbf{In turn, the energy of dissipation should be strongly dependent on the
structural factor (S): the ratio of kinetic energy of matter to its total
internal energy. We assume here that this dependence for viscosity calculation
is cubical: }$(T_{k}/U_{\text{tot}})^{3}=S^{3}$\textbf{.}

Consequently, the contributions of translational and librational
macrodeformons to resulting viscosity we present in the following way:%

\begin{equation}
\eta_{tr,lb}^{M}=\left[  {\frac{E_{D}^{M}}{\Delta v_{f}^{0}}}\cdot\tau
^{M}\left(  {\frac{T_{k}}{U_{\text{tot}}}}\right)  ^{3}\right]  _{tr,lb}%
\tag{45}%
\end{equation}

where: reduced fluctuating volume $(\Delta v_{f}^{0})$ corresponds to (44);
the energy of macrodeformons: $[E_{D}^{M}=-kT\cdot(\ln P_{D}^{M})]_{tr,lb}$.

The cycle-periods of the \textit{tr }and \textit{lib }macroeffectons has been
introduced as:%

\begin{equation}
\left[  \tau^{M}=\tau_{A}+\tau_{B}+\tau_{D}\right]  _{tr,lb}\tag{46}%
\end{equation}
where: characteristic life-times of macroeffectons in A, \thinspace B-states
and that of transition state in the volume of primary electromagnetic
deformons can be presented, correspondingly, as follows:%

\begin{equation}
\left[  \tau_{A}=\left(  \tau_{a}\cdot\tau_{\overline{a}}\right)
^{1/2}\right]  _{tr,lb}\text{ \ \ and\ \ \ \ }\left[  \tau_{A}=\left(
\tau_{a}\cdot\tau_{\overline{a}}\right)  ^{1/2}\right]  _{tr,lb}\tag{47}%
\end{equation}%

\[
\left[  \tau_{D}=\left|  (1/\tau_{A})-(1/\tau_{B})\right|  ^{-1}\right]
_{tr,lb}
\]

Using (47, 46 and 44) it is possible to calculate the contributions of
$\left(  A\rightleftharpoons B\right)  \,$ cycles of translational and
librational macroeffectons to viscosity separately, using (45).

The averaged contribution of macroexcitations (tr and lb)in viscosity is:%

\begin{equation}
\eta^{M}=\left[
\begin{array}
[c]{c}%
(\eta)_{tr}^{M}\cdot(\eta)_{lb}^{M}%
\end{array}
\right]  ^{1/2}\tag{48}%
\end{equation}

The contribution of \textit{a }and \textit{b convertons }to viscosity of
liquids could be presented in a similar to (44-48) manner after substituting
the parameters of \textit{tr and lb }macroeffectons with parameters of
\textit{a and b }convertons:%

\begin{equation}
\eta_{ac,bc}=\left[  {\frac{E_{c}}{\Delta v_{f}^{0}}}\tau_{c}\left(
{\frac{T_{k}}{U_{\text{tot}}}}\right)  ^{3}\right]  _{ac,bc}\tag{49}%
\end{equation}

where: reduced fluctuating volume of (\textit{a }and \textit{b}) convertons
$(\Delta v_{f}^{0})_{ac,bc}$ corresponds to:%

\begin{equation}
(\Delta v_{f}^{0})_{ac,bc}={\frac{V_{0}}{N_{0}n^{2}}}{\frac{1}{Z}}%
P_{ac,bc}\tag{50}%
\end{equation}

where: $P_{ac}$ and $P_{bc}$ are the relative probabilities of \textit{tr}%
/\textit{lib} interconversions between \textit{a} and \textit{b} states of
translational and librational primary effectons (see Introduction and Chapter
$4);\;E_{ac}$ and $E_{bc}$ are the excitation energies of (\textit{a} and
\textit{b}) convertons correspondingly (see Chapter 4 of [1] and [2]);

Characteristic life-times for \textit{ac}-convertons and \textit{bc}%
-convertons $[tr/lb]$ in the volume of primary librational effectons
(''flickering clusters'') could be presented as:%

\begin{equation}%
\begin{array}
[c]{l}%
\tau_{ac}=(\tau_{a})_{tr}+(\tau_{a})_{lb}=(1/\nu_{a})_{tr}+(1/\nu_{a})_{lb}\\
\tau_{bc}=(\tau_{b})_{tr}+(\tau_{b})_{lb}=(1/\nu_{b})_{tr}+(1/\nu_{b})_{lb}%
\end{array}
\tag{51}%
\end{equation}
The averaged contribution of the both types of convertons in viscosity is:%

\begin{equation}
\eta_{c}=(\eta_{ac}\cdot\eta_{bc})^{1/2}\tag{52}%
\end{equation}
This contribution could be responsible for microviscosity or better term:
\textbf{mesoviscosity}, related to volumes, equal to that of primary
librational effectons.

The resulting viscosity (Fig.5) is a sum of the averaged contributions of
macrodeformons and convertons:%

\begin{equation}
\eta=\eta^{M}+\eta_{c}\tag{53}%
\end{equation}

\begin{center}%
\begin{center}
\includegraphics[
height=3.736in,
width=4.3509in
]%
{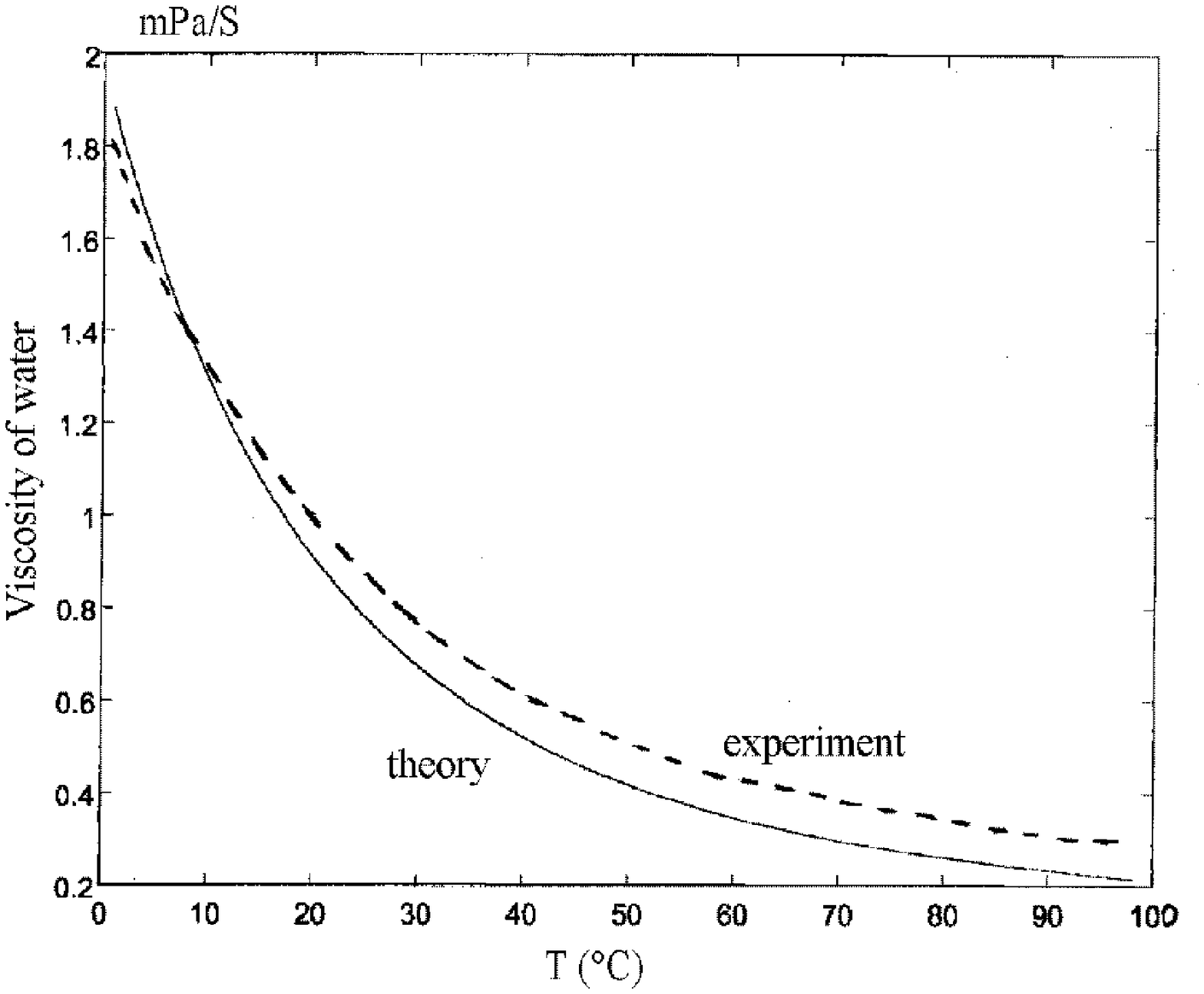}%
\end{center}
\medskip
\end{center}

\begin{quotation}
\textbf{Fig.~5. }Theoretical and experimental temperature dependences of
viscosities for water. Computer calculations were performed using eqs. (44 -
53) and (4.3; 4.36).
\end{quotation}

\medskip

The best correlation between theoretical and experimental data was achieved
after assuming that only ($\pi/2=2\pi/4)$ part of the period of above
described fluctuation cycles is important for dissipation and viscosity.
Introducing this factor to equations for viscosity calculations gives up very
good correspondence between theory and experiment in all temperature interval
(0-100$^{0}$C) for water (Fig.5).

As will be shown below the same factor, introducing the effective time of
fluctuations [$\frac\tau{\pi/2}$], leads to best results for self-diffusion
coefficient calculation.

In the classical hydrodynamic theory the sound absorption coefficient
($\alpha$) obtained by Stokes includes share $(\eta)$ and bulk $(\eta_{b})$
averaged macroviscosities:%

\begin{equation}
\alpha={\frac{\Omega}{2\rho v_{s}^{3}}}\left(  {\frac{4}{3}}\eta+\eta
_{b}\right)  ,\tag{54}%
\end{equation}
where $\Omega$ is the angular frequency of sound waves; $\rho$ is the density
of liquid.

Bulk viscosity ($\eta_{b}$) is usually calculated from the experimental $\eta$
and $\alpha$. It is known that for water:%

\[
(\eta_{b}/\eta)\sim3.
\]
\medskip

\smallskip

\begin{center}
{\large The viscosity of solids }
\end{center}

\smallskip

In accordance with our model, the biggest fluctuations: \textit{supereffectons
and superdeformons }(see Introduction) are responsible for viscosity and
diffusion phenomena in solid state. Superdeformons are accompanied by the
emergency of cavitational fluctuations in liquids and the defects in solids.
The presentation of viscosity formula in solids $(\eta_{s})$ is similar to
that for liquids:%

\begin{equation}
\eta_{S}={\frac{E_{S}}{(\Delta v_{f}^{0})_{S}}}\cdot\tau_{S}\left[
{\frac{T_{k}}{U_{\text{tot}}}}\right]  ^{3}\tag{55}%
\end{equation}

where: reduced fluctuating volume, related to superdeformons excitation
$(\Delta v_{f}^{0})_{s}$ is:%

\begin{equation}
(\Delta v_{f}^{0})_{S}={\frac{V_{0}}{N_{0}n^{2}}}{\frac{1}{Z}}P_{S}\tag{56}%
\end{equation}

where: $P_{s}=(P_{D}^{M})_{tr}\cdot(P_{D}^{M})_{lb}$ is the relative
probability of superdeformons, equal to product of probabilities of \textit{tr
and lb }macrodeformons excitation (see $42);\;E_{s}=-kT\cdot\ln P_{s}$ is the
energy of superdeformons (see Chapter 4);

Characteristic cycle-period of $(A^{*}\rightleftharpoons B^{*})$ transition of
supereffectons is related to its life-times in A$^{*},\;$B$^{*}$and transition
D$^{*}\;$states (see eq.46) as was shown in section 4.3:%

\begin{equation}
\tau_{S}=\tau_{A^{*}}+\tau_{B^{*}}+\tau_{D^{*}}\tag{56a}%
\end{equation}
The viscosity of ice, calculated from eq.(55) is bigger than that of water
(eq.53) to about $10^{5}$ times. This result is in accordance with available
experimental data.

\medskip

\begin{center}
{\Large 7. Brownian diffusion}
\end{center}

\smallskip

The important formula obtained by Einstein in his theory of Brownian motion is
for translational motion of particle:%

\begin{equation}
r^{2}=6Dt={\frac{kT}{\pi\eta a}}t\tag{57}%
\end{equation}
and that for rotational Brownian motion:%

\begin{equation}
\varphi^{2}={\frac{kT}{4\pi\eta a^{3}}}t\tag{58}%
\end{equation}
where: \textit{a }- radius of spherical particle, much larger than dimension
of molecules of liquid. The coefficient of diffusion D for Brownian motion is
equal to:%

\begin{equation}
D={\frac{kT}{6\pi\eta a}}\tag{59}%
\end{equation}
If we take the angle\thinspace$\bar{\varphi}^{2}=1/3\;$ in\ (59), then the
corresponding rotational correlation time comes to the form of the known
Stokes- Einstein equation:%

\begin{equation}
\tau={\frac{4}{3}}\pi a^{3}{\frac{1}{k}}\left(  {\frac{\eta}{T}}\right)
\tag{60}%
\end{equation}
All these formulas (57 - 60) include macroscopic share viscosity $(\eta)$
corresponding to our (53).

\medskip

\begin{center}
{\Large 8. Self-diffusion in liquids and solids}
\end{center}

\smallskip

Molecular theory of self-diffusion, as well as general concept of
\textit{transfer phenomena }in condensed matter is extremely important, but
still unresolved problem.

Simple semiempirical approach developed by Frenkel leads to following
expression for diffusion coefficient in liquid and solid:%

\begin{equation}
D={\frac{a^{2}}{\tau_{0}}}\exp(-W/kT)\tag{61}%
\end{equation}
where [a] is the distance of fluctuation jump; $\tau_{0}\sim(10^{-12}%
\div10^{-13})\,s$ is the average period of molecule oscillations between
jumps; W - activation energy of jump.

The parameters: a, $\tau_{0}$ and W should be considered as a fit parameters.

In accordance with \textbf{mesoscopic theory}, the process \textbf{of
self-diffusion }in liquids\textbf{, }like that of \textbf{viscosity},
described above, is determined by two contributions:

a)~the \textbf{collective, nonlocal contribution}, related to translational
and librational macrodeformons $(D_{tr,lb})$;

b)~the \textbf{local contribution, }related to coherent clusters flickering:
[dissociation/association] of primary librational effectons (\textit{a }and
\textit{b})- convertons $(D_{ac,bc})$.

\smallskip

Each component of the resulting coefficient of self-diffusion (D) in liquid
could be presented as the ratio of fluctuation volume cross-section surface:
$[\Delta v_{f}^{0}]^{2/3}$ to the period of macrofluctuation $(\tau)$. The
first contribution to coefficient \textbf{D,} produced by translational and
librational macrodeformons is:%

\begin{equation}
D_{tr,lb}=\left[  \left(  \Delta v_{f}^{0}\right)  ^{2/3}\cdot{\frac{1}%
{\tau^{M}}}\right]  _{tr,lb}\tag{62}%
\end{equation}

where: the surface cross-sections of reduced fluctuating free volumes (see
eq.43) fluctuations in composition of macrodeformons (\textit{tr \ and\ lb) }are:%

\begin{equation}
(\Delta v_{f}^{0})_{tr,lb}^{2/3}=\left[  {\frac{V_{0}}{N_{0}n^{2}}}{\frac
{1}{Z}}\exp\left(  -{\frac{\epsilon_{D}^{M}}{kT}}\right)  _{tr,lb}\right]
^{2/3}\tag{63}%
\end{equation}
\medskip($\tau^{M}$)$_{tr,lb}$ are the characteristic $(A\Leftrightarrow B) $
cycle-periods of translational and librational macroeffectons (see eqs. 46 and 47).

The averaged component of self-diffusion coefficient, which takes into account
both types of nonlocal fluctuations, related to translational and librational
macroeffectons and macrodeformons, can be find as:%

\begin{equation}
D^{M}=[(D)_{tr}^{M}\cdot(D)_{lb}^{M}]^{1/2}\tag{64}%
\end{equation}

\textbf{The formulae for the second, local contribution to self-diffusion }in
liquids, related to (\textit{a }and \textit{b}) convertons $(D_{ac,bc})$ are
symmetrical by form to that, presented above for nonlocal processes:%

\begin{equation}
D_{ac,bc}=\left[  (\Delta v_{f}^{0})^{2/3}\cdot{\frac{1}{\tau_{S}}}\right]
_{ac,bc}\tag{65}%
\end{equation}

where: reduced fluctuating free volume of (\textit{a }and \textit{b})
convertons $(\Delta v_{f}^{0})_{ac,bc}$ is the same as was used above in
mesoscopic theory of viscosity (eq.50):%

\begin{equation}
(\Delta v_{f}^{0})_{ac,bc}={\frac{V_{0}}{N_{0}n^{2}}}{\frac{1}{Z}}%
P_{ac,bc}\tag{66}%
\end{equation}

where:$\;P_{ac}$ and $P_{bc}$ are the relative probabilities of \textit{tr}%
/\textit{lib} interconversions between \textit{a} and \textit{b} states of
translational and librational primary effectons (see Introduction and Chapter 4)

The averaged local component of self-diffusion coefficient, which takes into
account both types of convertons (ac and bc) is:%

\begin{equation}
D_{C}=[(D)_{ac}\cdot(D)_{bc}]^{1/2}\tag{67}%
\end{equation}

In similar way we should take into account the contribution of macroconvertons
$(D_{Mc})$:%

\begin{equation}
D_{Mc}=\left(  \frac{V_{0}}{N_{0}n^{2}}\frac{1}{Z}P_{Mc}\right)  ^{2/3}%
\cdot\frac{1}{\tau_{Mc}}\tag{67a}%
\end{equation}

where: $P_{Mc}=P_{ac}\cdot P_{bc}$ is a probability of macroconverton excitation;

the life-time of macroconverton is:%

\begin{equation}
\tau_{Mc}=(\tau_{ac}\cdot\tau_{bc})^{1/2}\tag{67b}%
\end{equation}

The cycle-period of $(ac)$ and $(bc)$ convertons are determined by the sum of
life-times of intermediate states of primary translational and librational effectons:%

\begin{equation}
\tau_{ac}=(\tau_{a})_{tr}+(\tau_{a})_{lb};\;\text{ \ and\ \ \ \ }\tau
_{bc}=(\tau_{b})_{tr}+(\tau_{b})_{lb}\tag{67c}%
\end{equation}

The\textit{\ }life-times of primary and secondary effectons (lb and tr) in
\textit{a}- and \textit{b}-states are the reciprocal values of corresponding
state frequencies:%

\begin{equation}
\text{\lbrack}\tau_{a}=1/\nu_{a};\text{ \thinspace}\tau_{\overline{a}}%
=1/\nu_{\overline{a}};\;\text{ and\ \ \ }\tau_{b}=1/\nu_{b};\text{ \ }%
\tau\overline{_{b}}=1/\nu_{\overline{b}}\text{]}_{tr,lb}\tag{67d}%
\end{equation}
[$\nu_{a}$ and $\nu_{b}$]$_{tr,lb}$ correspond to eqs. 4.8 and 4.9;\ \ [$\nu
_{\overline{a}}$ and $\nu_{\overline{b}}$]$_{tr,lb}$\text{ }could be
calculated using eqs.2.54 and 2.55.

The resulting coefficient of self-diffusion in liquids (D) is a sum of
nonlocal $(D^{M})$ and local $(D_{c},\;D_{Mc})$ effects contributions (see
eqs.64 and 67):%

\begin{equation}
D=D^{M}+D_{c}+D_{Mc}\tag{68}%
\end{equation}
The effective fluctuation-times were taken the same as in previous section for
viscosity calculation, using the correction factor [($\pi/2)\cdot\tau] $.

\begin{center}%
\begin{center}
\includegraphics[
height=2.5149in,
width=5.578in
]%
{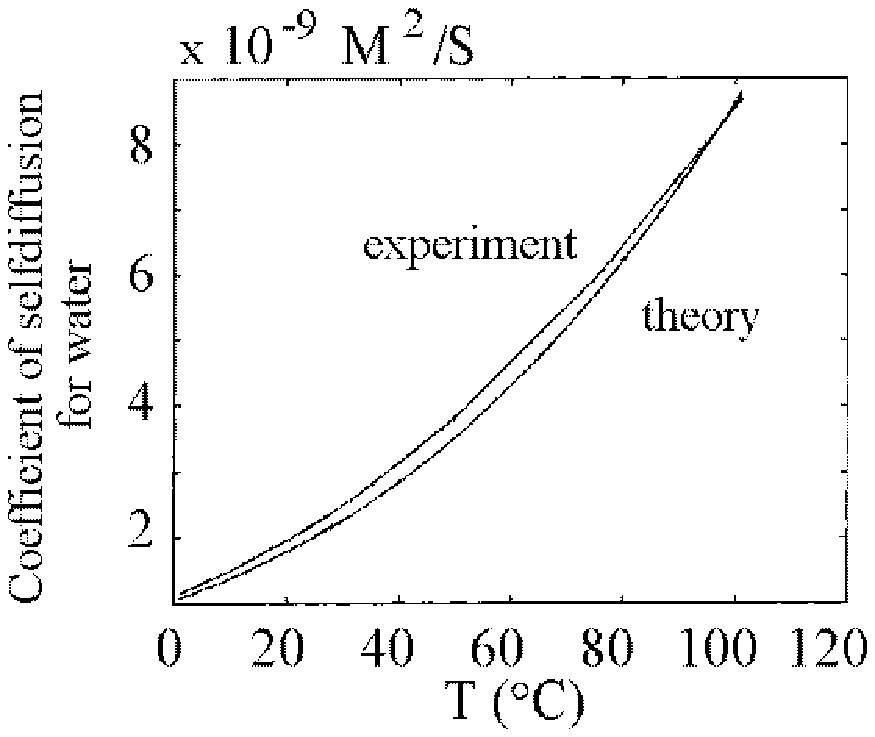}%
\end{center}
\medskip
\end{center}

\begin{quotation}
\textbf{Fig.~6. }Theoretical and experimental temperature dependences of
self-diffusion coefficients in water. Theoretical coefficient was calculated
using eq. 68.
\end{quotation}

\medskip

Like in the cases of thermal conductivity, viscosity and vapor pressure, the
results of theoretical calculations of self-diffusion coefficient coincide
well with experimental data for water (Fig. 6) in temperature interval
$(0-100^{0}C)$.

{\large \smallskip}

\begin{center}
{\large The self-diffusion in solids }
\end{center}

\smallskip

In solid state only the biggest fluctuations: \textit{superdeformons,
}representing simultaneous excitation of translational and librational
macrodeformons in the same volumes of matter are responsible for diffusion and
the viscosity phenomena. They are related to origination and migration of the
defects in solids. The formal presentation of superdeformons contribution to
self-diffusion in solids $(D_{s})$ is similar to that of macrodeformons for liquids:%

\begin{equation}
D_{S}=(\Delta v_{f}^{0})_{S}^{2/3}\cdot{\frac{1}{\tau_{S}}}\tag{69}%
\end{equation}

where: reduced fluctuating free volume in composition of superdeformons
$(\Delta v_{f}^{0})_{S}$ is the same as was used above in mesoscopic theory of
viscosity (eq.56):%

\begin{equation}
(\Delta v_{f}^{0})_{S}={\frac{V_{0}}{N_{0}n^{2}}}{\frac{1}{Z}}P_{S}\tag{70}%
\end{equation}

where: $P_{S}=(P_{D}^{M})_{tr}\cdot(P_{D}^{M})_{lb}$ \ is the relative
probability of superdeformons, equal to product of probabilities of \textit{tr
and lb }macrodeformons excitation (see 42).

Characteristic cycle-period of supereffectons is related to that of \textit{tr
and lb }macroeffectons like it was presented in eq.(56a):%

\begin{equation}
\tau_{s}=\tau_{A^{*}}+\tau_{B^{*}}+\tau_{D^{*}}\tag{71}%
\end{equation}

The self-diffusion coefficient for ice, calculated from eq.69 is less than
that of water (eq.53) to about $10^{5}$ times. This result is in accordance
with available experimental data.

Strong decreasing of D in a course of phase transition: [water $\rightarrow$
ice] predicted by our mesoscopic theory also is in accordance with experiment
(Fig. 7).

\begin{center}%
\begin{center}
\includegraphics[
height=2.6342in,
width=4.1537in
]%
{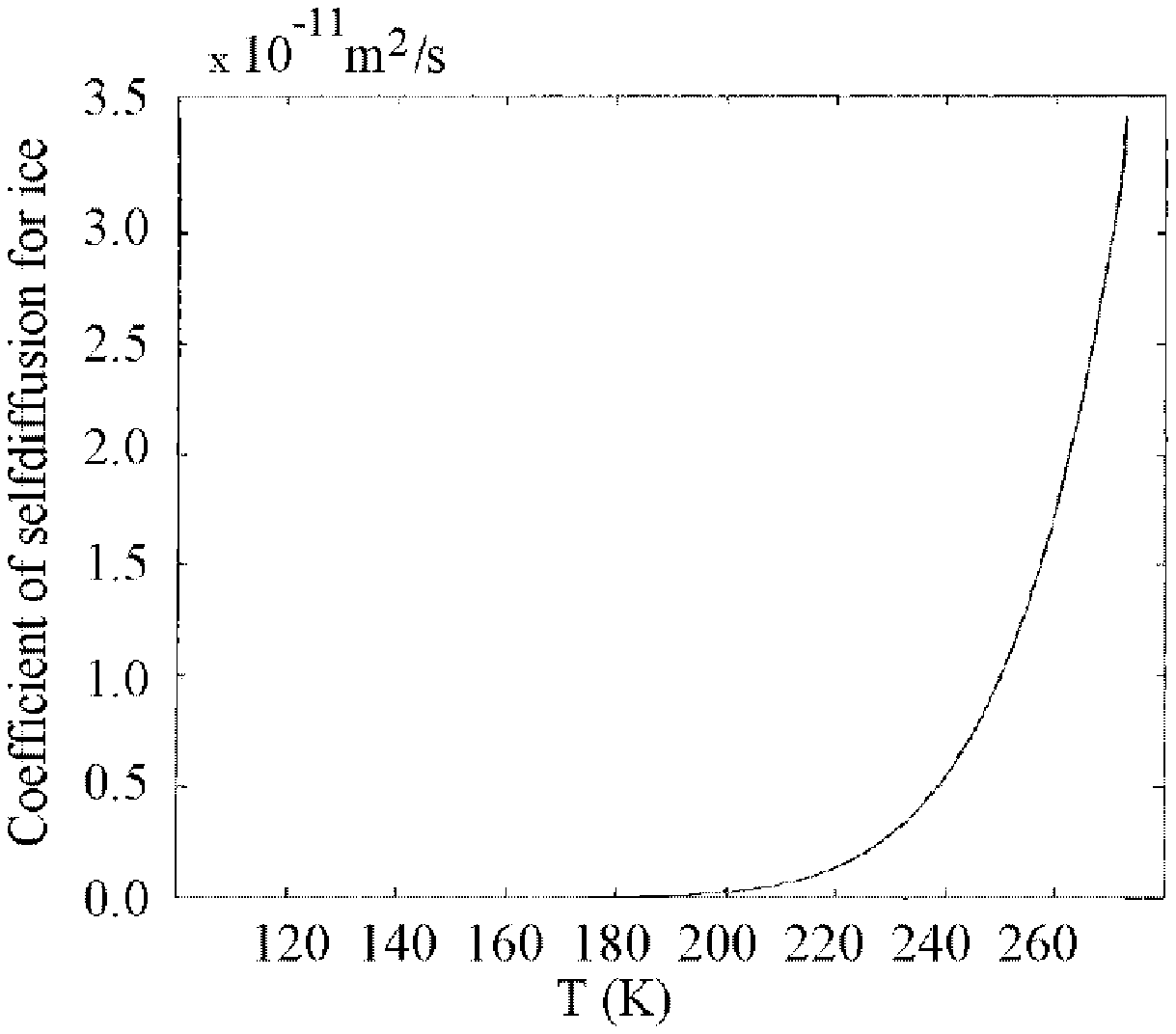}%
\end{center}
\medskip

Fig.~7. Theoretical temperature dependences of self-diffusion coefficients in ice.

\textbf{\smallskip}
\end{center}

\textbf{All these results allow to consider our mesoscopic theory of transfer
phenomena as a quantitatively confirmed one. They point that the ''mesoscopic
bridge'' between Micro- and Macro Worlds is wide and reliable indeed. It gives
a new possibilities for understanding and detailed description of very
different phenomena in solids and liquids.}

\textbf{\smallskip}

\textbf{One of the important consequences of our theory of viscosity and
diffusion is the possibility of explaining numerous nonmonotonic temperature
changes, registered by a number of physicochemical methods in various aqueous
systems during the study of temperature dependences ([6], [7], [8], [9]; [10];
[11], }$[12]$\textbf{).}

Most of them are related to diffusion or viscosity processes and may be
explained by nonmonotonic changes of the refraction index, included in our
equations: 44, 45, 50 for viscosity and eqs. 69, 70 for self-diffusion. For
water these temperature anomalies of refraction index were revealed
experimentally, using few wave lengths in the temperature interval
$3-95^{0}\;$[13]. They are close to Drost-Hansen temperatures. The explanation
of these effects, related to periodic variation of primary librational
effectons stability with monotonic temperature change was presented as
comments to Fig. 7a of [1] or Fig.4a of [2].

\bigskip

\textbf{Another consequence of our theory is the elucidation of a big
difference between librational }$\eta_{lb}\;(48)$\textbf{, translational
}$\eta_{tr}\;(45)$\textbf{\ viscosities and mesoviscosity, determined by
}$[lb/tr]$\textbf{\ convertons (49 and 52).}

\textbf{The effect of mesoviscosity can be checked as long as the volume of a
Brownian particle does not exceed much the volume of primary librational
effectons (eq. 15). If we take a Brownian particle, much bigger than the
librational primary effecton, then its motion will reflect only averaged share
viscosity (eq.53).}

\smallskip

\textbf{The third consequence of the mesoscopic theory of viscosity is the
prediction of nonmonotonic temperature behavior of the sound absorption
coefficient }$\alpha$\textbf{\ (51)}.\textbf{\ Its temperature dependence must
have anomalies in the same regions, where the refraction index has.}

\textbf{The experimentally revealed temperature anomalies of (n) also follow
from our theory as a result of nonmonotonic }$(a\Leftrightarrow b)_{lb}%
$\textbf{\ equilibrium\ behavior, stability of primary lb effectons and
probability of [lb/tr] convertons excitation (see Discussion to Fig.7a of [1]
or to Fig.4a of [2]).}

\smallskip

\textbf{Our model predicts also that in the} \textbf{course of transition from
the laminar type of flow to the turbulent one the share viscosity (}$\eta
$\textbf{) will increases due to increasing of structural factor }%
$(T_{k}/U_{\text{tot}})$\textbf{\ in eq. 45.}

\smallskip

\textbf{The superfluidity }$(\eta\rightarrow0)$\textbf{\ in the liquid helium
could be a result of inability of this liquid at the very low temperature for
translational and librational macroeffectons excitations, i.e. }$\tau
^{M}\rightarrow0$\textbf{.}

\textbf{In turn, it is a consequence of tending to zero the life-times of
secondary effectons and deformons in eqs.(45), responsible for dissipation
processes, due to their Bose-condensation and transformation to primary ones
(see Chapter 12 of [1]). The polyeffectons, stabilized by Josephson's
junctions between primary effectons form the superfluid component of liquid helium.}

\medskip

\begin{center}
{\Large 9. Mesoscopic approach to proton conductivity in water, ice and other
systems, \thinspace containing hydrogen bonds}

\textbf{\smallskip}
\end{center}

\textbf{The numerous models of proton transitions in water and ice are usually
related to migration of two types of defects in the ideal Bernal-Fouler
structure [14]:}

\textbf{1.~Ionic defects originated as a result of }$2H_{2}O$%
\textbf{\ dissociation to hydroxonium and hydroxyl ions:}%

\[
2H_{2}O\Leftrightarrow H_{3}O^{+}+\text{OH}^{-}
\]

\textbf{2.~Orientational Bjerrum defects are subdivided to D (dopplet) and L
(leer) ones.}

\textbf{D-defect (positive) corresponds to situation, when 2 protons are
placed between two oxygen atoms, instead of the normal structure of hydrogen
bond: }$O\ldots H-O$\textbf{\ containing 1 proton.}

\textbf{L-defect (negative) corresponds to opposite anomaly, when even 1
proton between two oxygens is absent. Reorientation of dipole moment of
}$H_{2}O$\textbf{\ in the case of D- and L-defects leads to origination of charges:}%

\begin{equation}
q_{B}=q_{D}^{+}=\mid q_{L}^{-}\mid=0.38e\tag{72}%
\end{equation}
\textbf{The interrelation between the charge of electron (e), Bjerrum charge
}$(q_{B})$\textbf{\ and ionic charge }$(e_{I})\;($\textbf{Onsager, Dupius,
1962) is:}%

\begin{equation}
e=e_{I}+q_{B}\tag{73}%
\end{equation}

\textbf{The general approach to problem of proton transition takes into
account both types of defects: ionic and orientational. It was assumed that
orientational defects originate and annihilate in the process of continuous
migration of ions H}$^{+}$\textbf{\ and OH}$^{-}$\textbf{\ through the water
medium. Krjachko (1987) considers DL-pairs as a cooperative water cluster with
linear dimensions of about }$15\AA$\textbf{\ and with ''kink''. The Bjerrum's
DL-pair is a limit case of such model.}

The protons conductivity in water must decrease with temperature increasing
due to decreasing and disordering of water clusters and chains.

The \textit{kink-soliton }model of orientational defects migration along the
$H_{2}O$ chain was developed by Sergienko (1986). Mobility of ionic defects
exceeds the orientational ones about 10 times.

But it is important to point out that the strong experimental evidence
confirming the existence of just Bjerrum type orientational defects are still absent.

\textbf{Our mesoscopic model of proton diffusion in ice, water and other
hydrogen bonds containing systems includes following stages:}

\textbf{1. Ionization of water molecules in composition of superdeformons and
ionic defects origination;}

\textbf{2.~Bordering by }$H_{3}^{+}O$\textbf{\ and }$\,HO^{-}\,$\textbf{\ the
opposite surface-sides of primary librational effectons;}

\textbf{3.~Tunneling of proton through the volume of primary effecton as a
coherent water cluster (Bose-particle);}

\textbf{4.~Diffusion of ions }$H_{3}^{+}O$\textbf{\ and }$\,HO^{-}%
\;$\textbf{in the less ordered medium between primary effectons can be
realized in accordance with fluctuation mechanism described above in Section
8. The velocity of this stage is less than tunneling.}

\medskip

\textbf{Transitions of protons and hydroxyl groups can occur also due to
exchange processes (Antonchenko, 1991) like:}%

\begin{equation}
H_{3}^{+}O+H_{2}O\Leftrightarrow H_{2}O+H_{3}^{+}O\tag{74}%
\end{equation}%

\begin{equation}
H_{2}O+HO^{-}\Leftrightarrow\text{ }HO^{-}+H_{2}O\tag{75}%
\end{equation}
\textbf{The rate of ions transferring due to exchange is about 10 times more,
than diffusion velocity, but slower than that, determined by tunneling jumps.}

\textbf{5.~The orientational defects can originate as a result of }$H_{2}%
O$\textbf{\ molecules rearrangements and conversions between translational and
librational effectons in composition of superdeformons. Activation energy of
superdeformons and macroconvertons in water is 10.2 kcal/M and about }%
$\,12$\textbf{\ kcal/M in ice (see 6.12; 6.13). The additional activation
energy about 2-3 kcal/M is necessary for subsequent reorientation of
surrounding molecules (Bjerrum, 1951).}

\textbf{Like the ionic defects, positive (D) and negative (L) defects can form
a separated pairs on the opposite sides of primary effectons, approximated by
parallelepiped. Such pairs means the effectons polarization.}

\textbf{\smallskip}

Probability of $H^{+}\;or\;HO^{-}$ tunneling through the coherent cluster -
primary effecton in the (a)-state is higher than that in the (b)-state as far
(see 1.30-1.32 of [1]):%

\begin{equation}
\lbrack E_{a}=T_{\text{kin}}^{a}+V_{a}]<[E_{b}=T_{\text{kin}}^{b}%
+V_{b}]\tag{76}%
\end{equation}
where: $T_{\text{kin}}^{a}=T_{\text{kin}}^{b}$ are the kinetic energies of (a)
and (b) states; $E_{b}-E_{a}=V_{b}-V_{a}$. is the difference between total and
potential energies of these states.

\textbf{In accordance with known theory of\ tunneling, the probability of
passing the particle with mass }$(m)$\textbf{\ through the barrier with
wideness (a) and height }$(\epsilon)$\textbf{\ has a following dependence on
these parameters:}%

\begin{equation}
\mid\psi_{a}\mid\sim\exp\left(  -{\frac{a}{b}}\right)  =\exp\left(
-{\frac{a(2m\epsilon)^{1/2}}{\hbar}}\right) \tag{77}%
\end{equation}
\textbf{where:}%

\begin{equation}
b=\hbar/(2m\epsilon)^{1/2}\tag{78}%
\end{equation}
\textbf{is the effective wave function fading length.}

\textbf{Parameter (}$b$\textbf{) is similar to wave B most probable amplitude
}$(A_{B})$\textbf{\ with total energy\thinspace}$E_{B}=\epsilon\;$%
\textbf{\ (see eq. 2.22 of [1]):}%

\begin{equation}
b=A_{B}=\hbar/(2mE_{B})^{1/2}\tag{79}%
\end{equation}
\textbf{With temperature decreasing the }$(a\Leftrightarrow b)_{tr,lb}%
$\textbf{\ equilibrium of primary effectons shifts to the left:}%

\begin{equation}
K_{a}{\ }_{\Leftrightarrow}{\ }_{b}=(P_{a}/P_{b})\rightarrow\infty\tag{80}%
\end{equation}
\textbf{where }$P_{a}\rightarrow1\;$\textbf{and\ }$P_{b}\rightarrow
0$\textbf{\ are the thermoaccessibilities of (a) and (b) states of primary
effectons (see eqs. 4.10-4.12). The linear dimensions of primary effectons of
ice also tend to infinity at T}$\rightarrow0$\textbf{.}

\textbf{In water the tunneling stage of proton conductivity can be related to
primary librational effectons only and their role increase with temperature
decreasing. Dimensions of translational effectons in water does not exceed
that of one molecule as it leads from our computer calculations.}

\textbf{Increasing of protons conductivity in ice with respect to water, in
accordance with our model, is a consequence of participation of translational
primary effectons in tunneling of\ }$\left[  H^{+}\right]  \;$\textbf{besides
librational ones, as well as significant elevation of primary librational
effectons dimensions. Increasing of the total contribution of tunneling
process in protons migration in ice rise up their resulting transferring
velocity comparing to water.}

\medskip

\textbf{The external electric field induce:}

\textbf{a)~redistribution of positive and negative charges on the surface of
primary effectons determined by ionic defects and corresponding orientational defects;}

\textbf{b)~orientation of polarized primary effectons in field, making
quasi-continuous polyeffectons chains and that of the effectons orchestrated superclusters.}

\textbf{These effects create the conditions for relay mechanism of }$\left[
H^{+}\right]  $\textbf{\ and }$H_{3}^{+}O$\textbf{\ transmitting in the
direction of electric field and }$\left[  HO^{-}\right]  $\textbf{\ in the
opposite one. In accordance with our hierarchic model, the }$\left[
H^{+}\right]  $\textbf{\ transition mechanism includes the alternation of
tunneling, exchange and usual diffusion processes.}

\medskip

\begin{center}
{\Large 10. Regulation of pH and shining of water by electromagnetic and
acoustic fields}
\end{center}

{\Large \smallskip}

In accordance with our model, water dissociation reaction:
\[
H_{2}O\Leftrightarrow H^{+}+HO^{-}
\]
leading to \textit{increase of protons concentration }is dependent on
probability of $\left[  A_{S}^{*}\rightarrow B_{S}^{*}\right]  \,\,$%
transitions in supereffectons. This means that stimulation of $\;\left[
A_{S}^{*}\rightarrow B_{S}^{*}\right]  $ transitions (superdeformons)
\textbf{by ultrasound} with resonant frequencies, corresponding to frequency
of these transitions, should lead to decreasing of pH, i.e. to increasing the
concentration of protons $[H^{+}]$.

\textbf{The }$\left[  A_{S}\rightarrow B_{S}\right]  $\textbf{\ transitions of
supereffectons can be accompanied by origination of cavitational fluctuations
(cavitational microbubbles). The opposite }$\left[  B_{S}\rightarrow
A_{S}\right]  $\textbf{\ transitions are related to the collapse of these
microbubbles. As a result of this adiabatic process, water vapor in the
bubbles is heated up to }$4000-6000{\ }^{0}$\textbf{K. The usual energy of
superdeformons in water (Section 6.3):}%

\begin{equation}
\epsilon_{D}^{S}=10.2\text{ kcal}/M\simeq RT^{*}\tag{81}%
\end{equation}
\textbf{correspond to local temperature }$T^{*}\simeq5000{\ }^{0}$\textbf{K.
For the other hand it is known, that even }$2000{\ }^{0}K$\textbf{\ is enough
already for partial dissociation of water molecules inside bubbles (about
0.01\% of total amount of bubble water).}

The variable pressure (P), generated by ultrasound in liquid is dependent on
its intensity $(I,\,wt/cm^{2})$ like:%

\begin{equation}
P=(\rho v_{s}I)^{1/2}\cdot4.6\cdot10^{-3}(\text{atm)}\tag{82}%
\end{equation}
where $\rho$ is density of liquid; $\;v_{s}$ - sound velocity $(m/s)$.

$\left[  A_{S}^{*}\rightarrow B_{S}^{*}\right]  $ transitions and cavitational
bubbles origination can be stimulated also by IR radiation with frequency,
corresponding to the activation energy of corresponding big fluctuations,
described in mesoscopic theory by \textit{superdeformons }and
\textit{macroconvertons. }

In such a way, using IR radiation and ultrasound it is possible to regulate a
lot of different processes in aqueous systems, depending on pH and water activity.

\textbf{The increasing of ultrasound intensity leads to increased cavitational
bubble concentration. The dependence of the resonance cavity radius
}$(R_{\text{res}})$\textbf{\ on ultrasound frequency (f) can be approximately
expressed as:}%

\begin{equation}
R_{\text{res}}=3000/f\tag{83}%
\end{equation}
\textbf{At certain conditions the water placed in the ultrasound field, begins
to shine in the region: }$300-600nm$\textbf{\ [15]. This shining
(sonoluminescense) is a consequence of electronic excitation of water ions and
molecules in the volume of cavitational bubbles.}

\textbf{When the conditions of ultrasound standing wave exist, the number of
bubbles and intensity of sonoluminescense is maximal.}

The intensity of shining is nonmonotonicly dependent on temperature with
maxima around $15,\,30,\,45$ \thinspace and\thinspace$65^{0}\,$ [6]. This
temperature corresponds to extremes of stability of primary librational
effectons, related to the number of $H_{2}O$ per effecton's edge $(\kappa)$
(see comments to Fig. 7a of [1] or to Fig 4a of [2]). An increase of inorganic
ion concentration, destabilizing (a)-state of these effectons, elevate the
probability of superdeformons and consequently, shining intensity.

The most probable reason of photon radiation is recombination of water
molecules, turning it into exited state:%

\begin{equation}
^{-}OH\text{ }+H^{+}\rightleftharpoons H_{2}O^{*}\rightarrow H_{2}O+h\nu
_{p}\tag{84}%
\end{equation}
Very different chemical reactions can be stimulated in the volume of
cavitational fluctuation by the external fields. The optimal resonant
parameters of these fields could be calculated using hierarchic theory.

\textbf{We propose here that the reaction of water molecules recombination
(84) could be responsible for coherent ''biophotons'' radiation by cell's and
microbes cultures and living organisms in visible and ultraviolet (UV) range.
The advances in biophoton research are described by Popp et al., 1992 [16]. }

\textbf{In accordance to our model, the cell's body filaments - microtubules
(MTs) ''catastrophe'' (cooperative reversible disassembly of MTs) is a result
of the internal water cavitational fluctuations due to superdeformons
excitation. Such collective process should be accompanied by dissociation and
recombination (84) of part of water molecules, localized in the hollow core of
microtubules, leading to high-frequency electromagnetic radiation (see:
http://arXiv.org/abs/physics/0003045). The coherent biophotons in the infrared
(IR) range are a consequence of }$\,\left(  a\rightleftharpoons b\right)
_{tr,lb}\,\,$\textbf{transitions of the water primary effectons in microtubules.}

\textbf{\medskip}

{\large We can see that} {\large lot of well working new theoretical models
for different physical phenomena, based on our Hierarchic theory of condensed
matter, can be elaborated. It means that this theory may serve as new
convenient scientific language.\smallskip}

\bigskip

\begin{center}
=======================================================================\bigskip
\end{center}

\begin{quotation}
\textbf{REFERENCES}

\medskip

\textbf{[1]. Kaivarainen A. Hierarchic Concept of Matter and Field. Water,
biosystems and elementary particles. New York, NY,1995, pp. 485.}

\textbf{[2]. \thinspace Kaivarainen A. New Hierarchic Theory of Matter General
for Liquids and Solids: dynamics, thermodynamics and mesoscopic structure of
water and ice }

\textbf{(see URL: http://www.karelia.ru/\symbol{126}alexk) and:}

\textbf{[3].} \textbf{Kaivarainen A. Hierarchic Concept of Condensed Matter
and its Interaction with Light: New Theories of Light Refraction, Brillouin
Scattering\ and M\"{o}ssbauer effect }

\textbf{(see URL: http://www.karelia.ru/\symbol{126}alexk). }

\textbf{[4]. Blakemore J.S. Solid state physics. Cambridge University Press,
Cambridge, N.Y. e.a, 1985.}

\textbf{[5]. Dote J.L., Kivelson D., Schwartz H. J.Phys.Chem. 1981, 85, 2169.}

\textbf{[6]. Drost-Hansen W. In: Colloid and Interface Science. Ed. Kerker M.
Academic Press, New York, 1976, p.267.}

\textbf{[7]. Drost-Hansen W., Singleton J. Lin. Our aqueous heritage: evidence
for vicinal water in cells. In: Fundamentals of Medical Cell Biology, v.3A,
Chemistry of the living cell, JAI\ Press Inc.,1992, p.157-180.}

\textbf{[8]. Johri G.K., Roberts J.A. Study of the dielectric response of
water using a resonant microwave cavity as a probe. J.Phys.Chem.
}$\,1990,\,94,7386 $\textbf{.}

\textbf{[9]. Aksnes G., Asaad A.N. Influence of the water structure on
chemical reactions in water. A study of proton-catalyzed acetal hydrolysis.
Acta Chem. Scand. }$1989,43,726-734$\textbf{.}

\textbf{[10]. Aksnes G., Libnau O. Temperature dependence of esther hydrolysis
in water. Acta Chem.Scand. }$1991,45,463-467$\textbf{.}

\textbf{[11]. K\"{a}iv\"{a}r\"{a}inen A.I. Solvent-dependent flexibility of
proteins and principles of their function. D.Reidel Publ.Co., Dordrecht,
Boston, Lancaster, 1985,\thinspace pp.290.}

\textbf{[12]. K\"{a}iv\"{a}r\"{a}inen A., Fradkova L., Korpela T. Separate
contributions of large- and small-scale dynamics to the heat capacity of
proteins. A new viscosity approach. Acta Chem.Scand. }$1993,47,456-460$\textbf{.}

\textbf{[13]. Frontas'ev V.P., Schreiber L.S. J. Struct. Chem. (USSR}%
$)\,$\textbf{6(1966)512}$.$

\textbf{[14]. Antonchenko V.Ya. Physics of water. Naukova dumka, Kiev, 1986.}

\textbf{[15]. Guravlev A.I. and Akopjan V.B. Ultrasound shining. Nauka,
Moscow, 1977.}

\textbf{[16]. Popp F.A., Li K.H. and Gu Q. Recent advances in biophoton
research. Singapore: World Scientific, 1992.}
\end{quotation}
\end{document}